\documentclass[oneside]{mscs} 
\usepackage[utf8x]{inputenc}
\usepackage[T1]{fontenc}
\usepackage[english]{babel}
\usepackage{pdfsync}
\let\proof\relax

\usepackage{amsmath,amsthm,amssymb,cmll,bussproofs}
\usepackage{tikz}
\usepackage{subfigure}
\usepackage{fouriernc}
\usepackage[autostyle=true, english=british, strict=true, autopunct=true]{csquotes}
\usepackage{multirow} 
\usepackage{blkarray}
\usepackage{xspace}
\usepackage{graphicx}
\usepackage{enumitem}
\usepackage{hyperref}
\hypersetup{
	pdfdisplaydoctitle=true,
	pdfinfo={
		Author={Clément Aubert, Thomas Seiller},
		Title={Characterizing co-NL by a Group Action},
		Subject={Linear Logic, Complexity, Geometry of Interaction, Pointer Machine},
		Keywords={Theoretical computer Science, Linear Logic, Implicit complexity, Proof nets, Operator algebra},
		Creator=	{PDFLaTeX},
		Producer={PDFLaTeX},
		Licence={Creative Commons BY-NC-SA3.0 or later},
		Lang={en}
	}
}

\theoremstyle{plain}
\newtheorem{theorem}{Theorem}
\newtheorem{corollary}[theorem]{Corollary}
\newtheorem{lemma}[theorem]{Lemma}
\newtheorem{proposition}[theorem]{Proposition}

\theoremstyle{definition}
\newtheorem{definition}[theorem]{Definition}

\theoremstyle{remark}

\tikzset{
>=latex, 
inner sep=0pt,
outer sep=2pt,
}
\tikzset{every picture/.style={line width=0.6pt}, rounded corners=3pt}

\usepackage{ulem}
\usepackage[agsm]{harvard}

\RequirePackage{filecontents}

\newcommand{\naturalN}{\mathbf{N}} 	
\newcommand{\complexN}{\mathbf{C}}	

\newcommand{\norm}[1]{\mathopen{\lVert}#1\mathclose{\rVert}} 
\newcommand{\hil}[1]{\mathbb{#1}} 
\newcommand{\vn}[1]{\mathfrak{#1}} 
\newcommand{\finhyp}{\vn{R}} 
\newcommand{\<}{\mathopen{<}} 
\renewcommand{\>}{\mathclose{>}} 
\newcommand{\B}[1]{\mathcal{L}(#1)} 

\newcommand{\N}[1]{\mathfrak{N}(#1)} 

\newcommand{\cc}[1]{\textbf{#1}} 

\newcommand{\sizeof}[1]{|#1|}
\newcommand{\bigO}{\mathcal{O}} 

\newcommand{\pb}[1]{\textbf{#1}\xspace} 
\newcommand{\stconncomp}{\pb{STConnComp}} 
\newcommand{\stconn}{\pb{STConn}} 

\newcommand{\lang}[1]{\ensuremath{\mathcal{L}(#1)}\xspace} 

\newcommand{\NDPM}{\ensuremath{\emph{NDPM}}\xspace} 
\newcommand{\NDPMs}{\ensuremath{\emph{NDPM}s}\xspace}
\newcommand{\NDPMp}[1]{\ensuremath{\emph{NDPM}(#1)}\xspace}

\newcommand{\pos}[1]{\sharp #1}

\author[C. Aubert and T. Seiller]
{C\ls L\ls É\ls M\ls E\ls N\ls T\ns A\ls U\ls B\ls E\ls R\ls T$^{\dagger}$\ns and\ns T\ls H\ls O\ls M\ls A\ls S\ns S\ls E\ls I\ls L\ls L\ls E\ls R$^{\ddagger}$ \thanks{This work was partly supported by the ANR-10-BLAN-0213 Logoi, the ANR-08-BLAN-0211-01 Complice and the GDR-IM's \enquote{visiting PhD student} Program.}\\
$\dagger$ Université Paris 13, Sorbonne Paris Cité, LIPN, CNRS, (UMR 7030), F-93430, Villetaneuse, France\\ \href{mailto:aubert@lipn.fr}{aubert@lipn.fr}
\addressbreak 
\\
$\ddagger$ I.H.\'{E}.S., Le Bois-Marie, 35, Route de Chartres, 91440 Bures-sur-Yvette, France\\ \href{mailto:seiller@ihes.fr}{seiller@ihes.fr}
}

\title{Characterizing \cc{co-NL} by a group action}
\date{15 September 2012 ; Revised 15 November 2013}

\usepackage[yyyymmdd,hhmmss]{datetime}
\usepackage{fancybox, graphicx}
\begin{document}
%
%

\maketitle

\begin{abstract}
In a recent paper, Girard \citeyear{normativity} proposed to use his recent construction of a geometry of interaction in the hyperfinite factor \cite{goi5} in an innovative way to characterize complexity classes. We begin by giving a detailed explanation of both the choices and the motivations of Girard's definitions. We then provide a complete proof that the complexity class \cc{co-NL} can be characterized using this new approach. We introduce the nondeterministic pointer machine as a technical tool, a concrete model to compute algorithms.
\end{abstract}

\section{Introduction}\label{sec:intro}
Traditionally, the study of complexity relies on the definition of programs based on some abstract machines, such as Turing machines. In recent years, a new approach to complexity stemmed from the so-called proofs-as-program – or Curry–Howard – correspondence which allows to understand program execution as a cut-elimination procedure in logic. This correspondence naturally extends to quantitative approaches that made it possible to work on complexity with tools coming from logic. Due to its resource-awareness, linear logic (\textbf{LL}) is particularly suitable to treat computational questions, and many bridges have been built between complexity classes and this formalism. To name a few, elementary linear logic (\textbf{ELL}) \cite{danos01}, soft linear logic \cite{lafont04} and bounded linear logic \cite{dal09} characterize complexity classes, but only deterministic, sequential and equal to \textbf{P} (polytime) or above. New directions have recently been explored to characterize other complexity classes: \textbf{SBAL} \cite{schopp07} characterizes \textbf{L} (logarithmic space), boolean proof nets \cite{aubert11,terui04}, was the first success toward a characterization of parallel classes.

All those attempts belong to the field of implicit computational complexity (ICC). One of the main advantages of the ICC approach is that it does not refer to a particular model or an external measuring condition. We only have to consider language restrictions (for instance by limiting the primitive recursion) or to infer the complexity properties of a program, for instance with techniques like quasi-interpretations. Linear logic offers a particularly nice framework to study complexity questions since the decomposition of implication into a linear implication and a duplication modality allows some fine tuning of the rules that govern the latter. All the previously quoted attempts are \emph{implicit} characterization of complexity classes, as those logical system rest on the limitation of the computational power of \textbf{LL}. Next to the restrictions of recursion and the rewriting system with quasi-interpretation, this approach exhibits several interesting results as there is no need to perform the computation to know the space or time needed.

The \emph{geometry of interaction} program \cite{goi3} was introduced by Girard a few years after the introduction of \textbf{LL}. In a first approximation, it aims at giving an interpretation of proofs --or programs-- that accounts for the dynamics of cut-elimination, hence of computation. Since the introduction of this program Girard proposed several constructions\footnote{The interested reader can find a more unifying approach in the second author's \enquote{Interaction Graphs} \cite{seiller-goiadd}.} to fulfill this program \cite{goi1,goi3,goi5}. Due to the fact that they are centered around the notion of computation, these constructions are particularly adapted to study computational complexity \cite{baillotpedicini,lago05}.

The approach studied in this paper, which was proposed recently by Girard, differs from the previous works on complexity. Indeed, though it uses the tools of Girard's geometry of interaction in the hyperfinite factor \cite{goi5}, its relation to the latter is restricted to the representation of integers which is, in this particular setting, uniform\footnote{
All (size of) inputs are represented as object in a unique space, whereas the naive GoI interpretation of integers as matrices (see Section \ref{sec:bin_int}) would yield matrices of varying sizes, hence not all elements of a single algebra.}: each integer is represented as an operator $N_{n}$ in the hyperfinite type $\text{II}_{1}$ factor $\finhyp$. By using an operator-theoretic construction —the crossed product— it is possible to internalize some isomorphisms acting on $\finhyp$. These internalized isomorphisms can be understood as sort of \enquote{basic instructions} one can use to define a sort of abstract machine. Such an abstract machine is thus an operator constructed using these basic instructions: the operators in the algebra generated by the internalizations of the isomorphisms. One can then define the language accepted by such an operator $\phi$: the set of natural numbers such that the product $\phi N_{n}$ is nilpotent.
We will only refer to \textbf{co-NL} and won't use the famous result that it is equal to \textbf{NL} \cite{immerman1988nondeterministic,Szelepcsenyi1987}, because it is more natural to think of our framework as capturing \emph{complementary} of complexity classes.
 
In this paper, we present in detail a first result obtained from this approach: considering the group of finite permutations of the natural numbers, we can obtain a characterization of the complexity class \cc{co-NL}. To ease the presentation and proofs of the result, we will introduce \emph{non-deterministic pointer machines}, which are a new characterization of \cc{co-NL} in terms of abstract machines.

\subsection*{Outline}\label{subsec:outline}
We start (Section \ref{sec:bin_int}) by explaining in detail, with numerous examples, how the proofs representing binary integers are represented by graphs in the setting of Geometry of Interaction, graphs that can be then seen as matrices. Computation will then be represented by the computation of the iterated products of a matrix $PN$ where $N$ is a matrix representing an integer and $P$ is a matrix representing the program. The nilpotency of this product will represent the fact that the integer represented by $N$ is accepted by the program represented by $P$.

But the representation of an integer as a matrix is non-uniform: the size of the matrix depends on the integer considered. Since we want the representations of programs to be able to handle any size of input, we embed the matrices representing integers into the hyperfinite factor, whose definition is recalled in Section \ref{sec:vna}. Operators that represent programs are constructed from finite permutations, which can be internalized --represented as operators-- using the crossed product construction. This allows them to perform some very simple operations on the input. Namely, they will be able to cope with several copies of the input and to scan them independently. However, since the representation of integers in the hyperfinite factor is not unique, one needs the notion of normative pair (Subsection \ref{subsec:norm_pair}) to guarantee that a program is insensitive to the chosen representation of the integer.

We next introduce, in Section \ref{sec:ndpm}, a notion of abstract machines -- non-deterministic pointer machines (\NDPM) -- well suited to be represented by operators. 
This model is very close to multi-head finite automata, a classical characterization of logspace computation, but we begin this section by presenting its specificities.
We then prove that \NDPMs can recognize any set in \cc{co-NL} by providing an example of a \cc{co-NL}-complete problem solved by a \NDPM and a mechanism of reduction between problems.

We then define (Section \ref{sec:encodingndpm}) an encoding of \NDPMs as a certain kind of operators --named boolean operators, which proves that \cc{co-NL} is contained in the set of languages accepted by such operators. To show the converse, we first show that checking the nilpotency of a product $PN$ in the hyperfinite factor, where $P$ is a boolean operator and $N$ represents an integer, is equivalent to checking that a certain matrix is nilpotent (this rest on the quite technical Lemma \ref{finiteNL}). Finally, we can show that deciding if this matrix is nilpotent is in \cc{co-NL}. 

\section{Binary Integers}\label{sec:bin_int}
In this paper, we will be working with binary integers. In this section, we will explain how it is possible to represent these integers by matrices. As it turns out, representation by matrices is not satisfactory, and it will be necessary to represent integers by operators acting on an infinite-dimensional (separable) Hilbert space, as it will be done in Section \ref{sec:integers}. 

In intuitionistic logic, binary lists are typed with $\forall X ~ (X \Rightarrow X) \Rightarrow ((X \Rightarrow X) \Rightarrow (X \Rightarrow X))$. In \textbf{ELL}, the type of binary lists is:
\begin{equation*}
\forall X~\oc(X\multimap X)\multimap (\oc(X\multimap X)\multimap \oc(X\multimap X))
\end{equation*}
To a binary integer\footnote{As binary lists trivially represent binary integers, we may focus on binary integers for free.} corresponds a proof of the sequent $\vdash \wn(X\otimes X^{\simbot}),\wn(X\otimes X^{\simbot}),\oc(X\multimap X)$. One can easily read from a proof of this sequent the binary list it represents by looking at the occurences of contraction (and in some cases, weakening) rules.
We develop below three examples: the empty list $\star$, the lists $\star0$ and $\star110$. In these examples, we labeled the variables in order to distinguish between the different occurrences of the variable $X$. This is necessary because we need to keep track of which formulas are principal in the contraction rule. This distinction, which by the way appears in geometry of interaction, is crucial since it may be the only difference between the proofs corresponding to two different binary lists. For instance, without this information, the proofs representing $\star110$ and $\star010$ would be exactly the same\footnote{Or more precisely, they would be indistinguishable. This is consequence of the fact that the contraction rule in sequent calculus does not distinguish which formulas it is contracting, even though it should in order to be correctly defined. For instance, there are more than one proof of $\vdash A,A$ obtained from $\vdash A,A,A$ by means of a contraction rule, but even though these are different, they are indistinguishable with the usual syntax of sequent calculus.}.

To each sequent calculus proof, we associate a graph which represents the axiom links in the sequent calculus proof. The vertices are arranged as a table where the different occurrences of the variables in the conclusion are represented on a horizontal scale, and a number of \emph{slices} are represented on a vertical scale: the contraction is represented in geometry of interaction by a superimposition which is dealt with by introducing new copies of the occurrences using the notion of slices. In previous works \cite{seiller-goim,seiller-goiadd}, one of the authors showed how to obtain a combinatorial version of (a fragment) of Girard's geometry of interaction in the hyperfinite factor. Though the graphs shown here are more complex than the ones considered in these papers (in particular, the edges may go from one \emph{slice} to another), they correspond exactly to the representation of binary lists in Girard's framework\footnote{For more details, one may consult Seiller's PhD Thesis \citeyear{seiller-phd}.}.

\begin{itemize}
\item[$\bullet$]The proof representing the empty list $\star$ uses the weakening rule twice:
\begin{prooftree}
\AxiomC{}
\RightLabel{\tiny{ax}}
\UnaryInfC{\scriptsize{$\vdash X(S)^{\simbot},X(E)$}}
\RightLabel{\tiny{$\parr$}}
\UnaryInfC{\scriptsize{$\vdash X(S)\multimap X(E)$}}
\RightLabel{\tiny{$\oc$}}
\UnaryInfC{\scriptsize{$\vdash \oc(X(S)\multimap X(E))$}}
\RightLabel{\tiny{$\wn_{w}$}}
\UnaryInfC{\scriptsize{$\vdash \wn(X(0i)\otimes X(0o)^{\simbot}),\oc(X(S)\multimap X(E))$}}
\RightLabel{\tiny{$\wn_{w}$}}
\UnaryInfC{\scriptsize{$\vdash \wn(X(0i)\otimes X(0o)^{\simbot}), \wn(X(1i)\otimes X(1o)^{\simbot}),\oc(X(S)\multimap X(E))$}}
\RightLabel{\tiny{$\parr$}}
\UnaryInfC{\scriptsize{$\vdash \wn(X(0i)^{\simbot}\multimap X(0o)^{\simbot}), \wn(X(1i)\otimes X(1o)^{\simbot}),\oc(X(S)\multimap X(E))$}}
\RightLabel{\tiny{$\parr$}}
\UnaryInfC{\scriptsize{$\vdash \wn(X(0i)^{\simbot}\multimap X(0o)^{\simbot}), \wn(X(1i)^{\simbot}\multimap X(1o)^{\simbot}),\oc(X(S)\multimap X(E))$}}
\RightLabel{\tiny{$\parr$}}
\UnaryInfC{\scriptsize{$\vdash \wn(X(0i)^{\simbot}\multimap X(0o)^{\simbot}), (\oc(X(1i)\multimap X(1o)) \multimap(\oc(X(S)\multimap X(E))))$}}
\RightLabel{\tiny{$\parr$}}
\UnaryInfC{\scriptsize{$\vdash (\oc(X(0i)\multimap X(0o)) \multimap ((\oc(X(1i)\multimap X(1o)) \multimap(\oc(X(S)\multimap X(E)))))$}}
\RightLabel{\tiny{$\forall$}}
\UnaryInfC{\scriptsize{$\vdash \forall X~(\oc(X(0i)\multimap X(0o))\multimap (\oc(X(1i)\multimap X(1o))\multimap \oc(X(S)\multimap X(E))))$}}
\end{prooftree}
We will use a double line in the following to ignore the bureaucracy of introducing all the $\parr$.
The corresponding graph is:
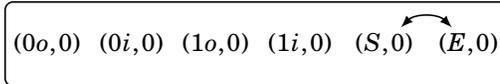
\begin{center}
\begin{tikzpicture}[x=0.7cm,y=0.7cm, scale=0.8]
	\node (Z00) at (0,-2) {$(0o,0)$};
	\node (Z01) at (2,-2) {$(0i,0)$};
	\node (Z10) at (4,-2) {$(1o,0)$};
	\node (Z11) at (6,-2) {$(1i,0)$};
	\node (ZD) at (8,-2) {$(S,0)$};
	\node (ZF) at (10,-2) {$(E,0)$};
	
	\draw[<->] (ZD) to [bend left=45] (ZF) {};

\draw[-] (-1,-3) rectangle (11,-1);
\end{tikzpicture}
\end{center}
\item[$\bullet$]The proof representing the list $\star0$ (resp. $\star1$) uses a weakening to introduce $X(1i)\multimap X(1o)$ (resp. $X(0i)\multimap X(0o)$):

\begin{prooftree}
\AxiomC{}
\RightLabel{\tiny{ax}}
\UnaryInfC{\scriptsize{$\vdash X(S)^{\simbot},X(0i)$}}
\AxiomC{}
\RightLabel{\tiny{ax}}
\UnaryInfC{\scriptsize{$\vdash X(0o)^{\simbot},X(E)$}}
\RightLabel{\tiny{$\otimes$}}
\BinaryInfC{\scriptsize{$\vdash X(0i)\otimes X(0o)^{\simbot},X(S)^{\simbot},X(E)$}}
\RightLabel{\tiny{$\parr$}}
\UnaryInfC{\scriptsize{$\vdash X(0i)\otimes X(0o)^{\simbot},X(S)\multimap X(E)$}}
\RightLabel{\tiny{$\oc$}}
\UnaryInfC{\scriptsize{$\vdash \wn(X(0i)\otimes X(0o)^{\simbot}),\oc(X(S)\multimap X(E))$}}
\RightLabel{\tiny{$\wn_{w}$}}
\UnaryInfC{\scriptsize{$\vdash \wn(X(0i)\otimes X(0o)^{\simbot}), \wn(X(1i)\otimes X(1o)^{\simbot}),\oc(X(S)\multimap X(E))$}}
\doubleLine
\UnaryInfC{\scriptsize{$\vdash \oc(X(0i)\multimap X(0o))\multimap (\oc(X(1i)\multimap X(1o))\multimap \oc(X(S)\multimap X(E)))$}}
\RightLabel{\tiny{$\forall$}}
\UnaryInfC{\scriptsize{$\vdash \forall X~\oc(X(0i)\multimap X(0o))\multimap (\oc(X(1i)\multimap X(1o))\multimap \oc(X(S)\multimap X(E)))$}}
\end{prooftree}

The corresponding graph is:
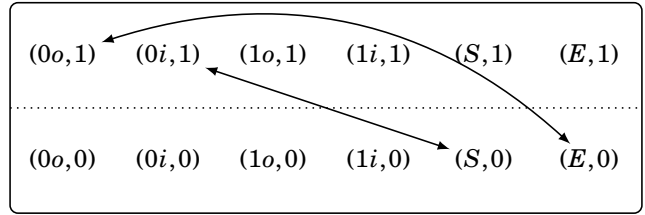
\begin{center}
\begin{tikzpicture}[x=0.7cm,y=0.7cm]
	\node (Z00) at (0,-2) {$(0o,0)$};
	\node (Z01) at (2,-2) {$(0i,0)$};
	\node (Z10) at (4,-2) {$(1o,0)$};
	\node (Z11) at (6,-2) {$(1i,0)$};
	\node (ZD) at (8,-2) {$(S,0)$};
	\node (ZF) at (10,-2) {$(E,0)$};

	\node (A00) at (0,0) {$(0o,1)$};
	\node (A01) at (2,0) {$(0i,1)$};
	\node (A10) at (4,0) {$(1o,1)$};
	\node (A11) at (6,0) {$(1i,1)$};
	\node (AD) at (8,0) {$(S,1)$};
	\node (AF) at (10,0) {$(E,1)$};

	\draw[<->] (A00) to [bend left](ZF) {};
	\draw[<->] (ZD) to (A01) {};

\draw[-] (-1,-3) rectangle (11,1);	
	\draw[-,dotted] (-1,-1) -- (11,-1) {};
\end{tikzpicture}
\end{center}

\item[$\bullet$]The proof representing the list $\star110$ contracts the occurrences $X(Ai)\otimes X(Ao)^{\simbot}$ and $X(1i)\otimes X(1o)^{\simbot}$, in bold below:

\begin{prooftree}
\AxiomC{}
\RightLabel{\tiny{ax}}
\UnaryInfC{\scriptsize{$\vdash X(0i),X(E)^{\simbot}$}}
\AxiomC{}
\RightLabel{\tiny{ax}}
\UnaryInfC{\scriptsize{$\vdash X(1i),X(0o)^{\simbot}$}}
\RightLabel{\tiny{$\otimes$}}
\BinaryInfC{\scriptsize{$\vdash X(0i)\otimes X(0o)^{\simbot},X(1i),X(E)^{\simbot}$}}
\AxiomC{}
\RightLabel{\tiny{ax}}
\UnaryInfC{\scriptsize{$\vdash X(Ai),X(1o)^{\simbot}$}}
\RightLabel{\tiny{$\otimes$}}
\BinaryInfC{\scriptsize{$\vdash X(0i)\otimes X(0o)^{\simbot},X(1i)\otimes X(1o)^{\simbot},X(Ai),X(E)^{\simbot}$}}
\AxiomC{}
\RightLabel{\tiny{ax}}
\UnaryInfC{\scriptsize{$\vdash X(S),X(Ao)^{\simbot}$}}
\RightLabel{\tiny{$\otimes$}}
\BinaryInfC{\scriptsize{$\vdash X(0i)\otimes X(0o)^{\simbot},X(1i)\otimes X(1o)^{\simbot},X(Ai)\otimes X(Ao)^{\simbot},X(S),X(E)^{\simbot}$}}
\RightLabel{\tiny{$\parr$}}
\UnaryInfC{\scriptsize{$\vdash X(0i)\otimes X(0o)^{\simbot},X(1i)\otimes X(1o)^{\simbot},X(Ai)\otimes X(Ao)^{\simbot},X(S)\multimap X(E)$}}
\RightLabel{\tiny{$\oc$}}
\UnaryInfC{\scriptsize{$\vdash \wn(X(0i)\otimes X(0o)^{\simbot}),\mathbf{\wn(X(1i)\otimes X(1o)^{\simbot}, \wn(X(Ai)\otimes X(Ao)^{\simbot}} , \oc(X(S)\multimap X(E))$}}
\RightLabel{\tiny{$\wn_{c}$}}
\UnaryInfC{\scriptsize{$\vdash \wn(X(0i)\otimes X(0o)^{\simbot}),\mathbf{\wn(X(1i)\otimes X(1o)^{\simbot})},\oc(X(S)\multimap X(E))$}}
\doubleLine
\UnaryInfC{\scriptsize{$\vdash \forall X~\oc(X(0i)\multimap X(0o))\multimap(\oc(X(1i)\multimap X(1o))\multimap \oc(X(S)\multimap X(E)))$}}
\end{prooftree}

The corresponding graph is:
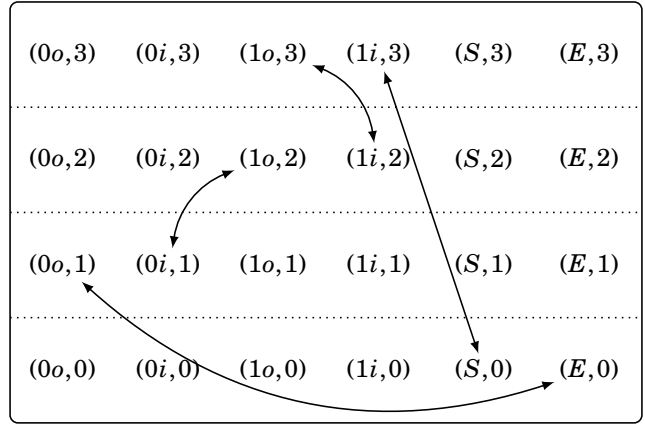
\begin{center}
\begin{tikzpicture}[x=0.7cm,y=0.7cm]
	\node (Z00) at (0,-2) {$(0o,0)$};
	\node (Z01) at (2,-2) {$(0i,0)$};
	\node (Z10) at (4,-2) {$(1o,0)$};
	\node (Z11) at (6,-2) {$(1i,0)$};
	\node (ZD) at (8,-2) {$(S,0)$};
	\node (ZF) at (10,-2) {$(E,0)$};

	\node (A00) at (0,0) {$(0o,1)$};
	\node (A01) at (2,0) {$(0i,1)$};
	\node (A10) at (4,0) {$(1o,1)$};
	\node (A11) at (6,0) {$(1i,1)$};
	\node (AD) at (8,0) {$(S,1)$};
	\node (AF) at (10,0) {$(E,1)$};

	\node (B00) at (0,2) {$(0o,2)$};
	\node (B01) at (2,2) {$(0i,2)$};
	\node (B10) at (4,2) {$(1o,2)$};
	\node (B11) at (6,2) {$(1i,2)$};
	\node (BD) at (8,2) {$(S,2)$};
	\node (BF) at (10,2) {$(E,2)$};

	\node (C00) at (0,4) {$(0o,3)$};
	\node (C01) at (2,4) {$(0i,3)$};
	\node (C10) at (4,4) {$(1o,3)$};
	\node (C11) at (6,4) {$(1i,3)$};
	\node (CD) at (8,4) {$(S,3)$};
	\node (CF) at (10,4) {$(E,3)$};
		
	\draw[<->] (A00) to [bend right] (ZF) {};
	\draw[<->] (A01) to [bend left] (B10) {};
	\draw[<->] (B11) to [bend right] (C10) {};
	\draw[<->] (ZD) to (C11) {};
	
\draw[-] (-1,-3) rectangle (11,5);	
	\draw[-,dotted] (-1,3) -- (11,3) {};
	\draw[-,dotted] (-1,1) -- (11,1) {};
	\draw[-,dotted] (-1,-1) -- (11,-1) {};
\end{tikzpicture}
\end{center}
\end{itemize}
The edges of the graph describe the scanning of the list. We illustrate this by explaining how to construct the graph corresponding to the list $\star 11010$. Indeed, the graph can be described directly from the list itself:
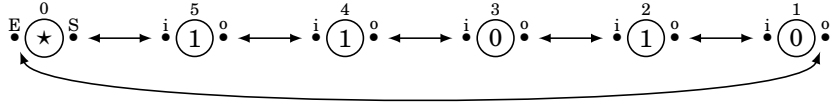
\label{expl_pare}

\begin{center}
\begin{tikzpicture}[inner sep=2pt, outer sep=1pt]
	\node[shape=circle,draw] (S) at (0,0) {$\star$};
		\node (SE) at (S.east) {};
		\node (SW) at (S.west) {};
		\node (SN) at (S.north) {};
		\node (So) at (SE.east) {$\bullet$};
		\node (SoL) at (So.north) {\scriptsize{S}};
		\node (Si) at (SW.west) {$\bullet$};
		\node (SiL) at (Si.north) {\scriptsize{E}};
		\node (St) at (SN.north) {\scriptsize{$0$}};
	\node[shape=circle,draw] (A) at (2,0) {$1$};
		\node (AE) at (A.east) {};
		\node (AW) at (A.west) {};
		\node (AN) at (A.north) {};
		\node (Ao) at (AE.east) {$\bullet$};
		\node (AoL) at (Ao.north) {\scriptsize{o}};
		\node (Ai) at (AW.west) {$\bullet$};
		\node (AiL) at (Ai.north) {\scriptsize{i}};
		\node (At) at (AN.north) {\scriptsize{$5$}};
	\node[shape=circle,draw] (B) at (4,0) {$1$};
		\node (BE) at (B.east) {};
		\node (BW) at (B.west) {};
		\node (BN) at (B.north) {};
		\node (Bo) at (BE.east) {$\bullet$};
		\node (BoL) at (Bo.north) {\scriptsize{o}};
		\node (Bi) at (BW.west) {$\bullet$};
		\node (BiL) at (Bi.north) {\scriptsize{i}};
		\node (Bt) at (BN.north) {\scriptsize{$4$}};
	\node[shape=circle,draw] (C) at (6,0) {$0$};
		\node (CE) at (C.east) {};
		\node (CW) at (C.west) {};
		\node (CN) at (C.north) {};
		\node (Co) at (CE.east) {$\bullet$};
		\node (CoL) at (Co.north) {\scriptsize{o}};
		\node (Ci) at (CW.west) {$\bullet$};
		\node (CiL) at (Ci.north) {\scriptsize{i}};
		\node (Ct) at (CN.north) {\scriptsize{$3$}};
	\node[shape=circle,draw] (D) at (8,0) {$1$};
		\node (DE) at (D.east) {};
		\node (DW) at (D.west) {};
		\node (DN) at (D.north) {};
		\node (Do) at (DE.east) {$\bullet$};
		\node (DoL) at (Do.north) {\scriptsize{o}};
		\node (Di) at (DW.west) {$\bullet$};
		\node (DiL) at (Di.north) {\scriptsize{i}};
		\node (Dt) at (DN.north) {\scriptsize{$2$}};
	\node[shape=circle,draw] (E) at (10,0) {$0$};
		\node (EE) at (E.east) {};
		\node (EW) at (E.west) {};
		\node (EN) at (E.north) {};
		\node (Eo) at (EE.east) {$\bullet$};
		\node (EoL) at (Eo.north) {\scriptsize{o}};
		\node (Ei) at (EW.west) {$\bullet$};
		\node (EiL) at (Ei.north) {\scriptsize{i}};
		\node (Et) at (EN.north) {\scriptsize{$1$}};
	\draw[<->] (So) to (Ai) {};
	\draw[<->] (Ao) to (Bi) {};
	\draw[<->] (Bo) to (Ci) {};
	\draw[<->] (Co) to (Di) {};
	\draw[<->] (Do) to (Ei) {};
	\draw[<->] (Eo) .. controls (10,-1) and (0,-1) .. (Si) {};
\end{tikzpicture}
\end{center}
Each element of the list lives in a different \textit{slice} —the integer shown above each element of the list. Moreover, each element is connected by its output vertex to its successor's input vertex (the successor of the last element is $\star$), and by its input vertex to its predecessor's output node.
This gives the following graph, which is the representation of $\star11010$:
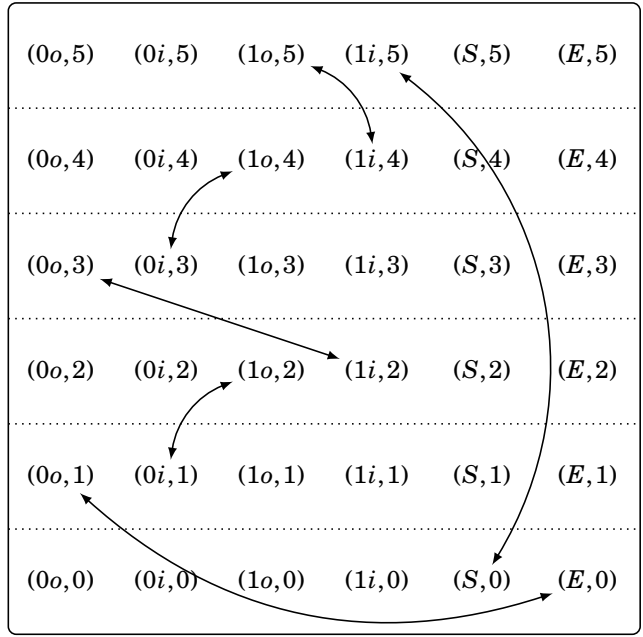
\begin{center}
\begin{tikzpicture}[x=0.7cm,y=0.7cm]
	\node (Z00) at (0,-2) {$(0o,0)$};
	\node (Z01) at (2,-2) {$(0i,0)$};
	\node (Z10) at (4,-2) {$(1o,0)$};
	\node (Z11) at (6,-2) {$(1i,0)$};
	\node (ZD) at (8,-2) {$(S,0)$};
	\node (ZF) at (10,-2) {$(E,0)$};

	\node (A00) at (0,0) {$(0o,1)$};
	\node (A01) at (2,0) {$(0i,1)$};
	\node (A10) at (4,0) {$(1o,1)$};
	\node (A11) at (6,0) {$(1i,1)$};
	\node (AD) at (8,0) {$(S,1)$};
	\node (AF) at (10,0) {$(E,1)$};

	\node (B00) at (0,2) {$(0o,2)$};
	\node (B01) at (2,2) {$(0i,2)$};
	\node (B10) at (4,2) {$(1o,2)$};
	\node (B11) at (6,2) {$(1i,2)$};
	\node (BD) at (8,2) {$(S,2)$};
	\node (BF) at (10,2) {$(E,2)$};

	\node (C00) at (0,4) {$(0o,3)$};
	\node (C01) at (2,4) {$(0i,3)$};
	\node (C10) at (4,4) {$(1o,3)$};
	\node (C11) at (6,4) {$(1i,3)$};
	\node (CD) at (8,4) {$(S,3)$};
	\node (CF) at (10,4) {$(E,3)$};

	\node (D00) at (0,6) {$(0o,4)$};
	\node (D01) at (2,6) {$(0i,4)$};
	\node (D10) at (4,6) {$(1o,4)$};
	\node (D11) at (6,6) {$(1i,4)$};
	\node (DD) at (8,6) {$(S,4)$};
	\node (DF) at (10,6) {$(E,4)$};
	
	\node (E00) at (0,8) {$(0o,5)$};
	\node (E01) at (2,8) {$(0i,5)$};
	\node (E10) at (4,8) {$(1o,5)$};
	\node (E11) at (6,8) {$(1i,5)$};
	\node (ED) at (8,8) {$(S,5)$};
	\node (EF) at (10,8) {$(E,5)$};

	\draw[<->] (A00) to [bend right] (ZF) {};
	\draw[<->] (A01) to [bend left] (B10) {};
	\draw[<->] (B11) to (C00) {};
	\draw[<->] (D10) to [bend right] (C01) {};
	\draw[<->] (D11) to [bend right] (E10) {};
	\draw[<->] (E11) to [bend left=42] (ZD) {};
	
\draw[-] (-1,-3) rectangle (11,9);
	\draw[-,dotted] (-1,-1) -- (11,-1) {};
	\draw[-,dotted] (-1,1) -- (11,1) {};
	\draw[-,dotted] (-1,3) -- (11,3) {};
	\draw[-,dotted] (-1,5) -- (11,5) {};
	\draw[-,dotted] (-1,7) -- (11,7) {};
\end{tikzpicture}
\end{center}
\begin{definition}[Matricial representation of a list]\label{def_matricial_rep_integer}
Given a binary representation of an integer $n=\star a_1, \hdots, a_k$ of size $k \neq 0$\footnote{We will always assume in the following that the length of the binary integer representing the integer $n$ under study is denoted by $k$.} and its corresponding graph $G_n$, $n$ is represented by $M_n$, a $6 \times 6$ block matrix of the following form:
\[
M_{n}=
\begin{blockarray}{ccccccc}
\BAmulticolumn{2}{c}{{\overbrace{~~~~~~~~~~~~~~}^0}} & \BAmulticolumn{2}{c}{{\overbrace{~~~~~~~~~~~~~~}^1}}& \BAmulticolumn{2}{c}{{\overbrace{~~~~~~~~~~~~~~}^{\ast}}} \\
\begin{block}{(cccccc)c}
	0 & l_{00} & 0 & l_{10} & s_{0} & 0 & \multirow{2}{*}{$\left.\begin{tabular}{c} \hspace{-0.5cm} \vspace{0.4cm} \end{tabular}\right\}\scriptstyle{0}$}\\
	l_{00}^{\ast} & 0 & l_{01}^{\ast} & 0 & 0 & e_{0}^{\ast}& \\
	0 & l_{01} & 0 & l_{11} & s_{1} & 0 & \multirow{2}{*}{$\left.\begin{tabular}{c} \hspace{-0.5cm} \vspace{0.4cm} \end{tabular}\right\} \scriptstyle{1}$}\\
	l_{10}^{\ast} & 0 & l_{11}^{\ast} & 0 & 0 & e_{1}^{\ast}& \\
	s_{0}^{\ast} & 0 & s_{1}^{\ast} & 0 & 0 & 0 & \multirow{2}{*}{$\left.\begin{tabular}{c} \hspace{-0.5cm} \vspace{0.4cm} \end{tabular}\right\} \scriptstyle{*}$}\\
	0 & e_{0} & 0 & e_{1} & 0 & 0 & \\
\end{block}
\end{blockarray}
\]
where coefficients are $(k+1) \times (k+1)$ matrices (the $(\cdot)^{\ast}$ denotes the conjugate-transpose) defined -- for $u, v \in \{0, 1\}$ -- by:
\begin{itemize}
\item $(l_{uv})_{a,b}=1$ if there is an edge in $G_n$ from $(uo, a)$ to $(vi, b)$, and $(l_{uv})_{a,b}=0$ otherwise;
\item $(e_u)_{0, n}=1$ if there is an edge from $(uo, n)$ to $(E, 0)$, and $(e_{u})_{a,b}=0$ otherwise;
\item $(s_v)_{0, n}=1$ if there is an edge from $(vo, n)$ to $(S, 0)$, and $(s_{v})_{a,b}=0$ otherwise.
\end{itemize}
One can simply make sure that no information is lost, the graph $G_n$ -- and by transitivity the input $n$ -- is totally and faithfully encoded in $M_n$.
\end{definition}

This representation of binary integers is however non-uniform: the size of the matrix depends on the size of the binary representation. This is where the use of von Neumann algebras takes its importance: any matrix algebra can be embedded in the type $\text{II}_{1}$ hyperfinite factor $\finhyp$. To get a uniform representation of integers, we therefore only need to embed the matricial representation in $\finhyp$. Before explaining this step in Section \ref{sec:integers}, we review in the next section some basics of the theory of von Neumann algebras. The aim of this section is not to introduce the reader to the theory which is much too rich to be condensed here, but to give some ideas and intuitions on it. In the end of the next section, we introduce the crossed product construction, an operation which will be fundamental in the subsequent sections.

\section{Von Neumann Algebras and Crossed Products}\label{sec:vna}

This section aims at giving a quick overview of the theory of von Neumann algebras. Most of the material it contains is not needed for understanding the results that follow, and the reader can skip this section for a first reading. Section \ref{sec:integers} uses the fact that we are working in the type $\text{II}_{1}$ hyperfinite factor, but the only results it uses is the fact that any matrix algebra can be embedded in a type $\text{II}_{1}$ factor (Proposition \ref{matrixintofactor}), the definition of the crossed product algebra (Definition \ref{crossedproduct}) and some properties of unitary operators acting on a Hilbert space. The remaining sections of the paper do not use results of the theory of operator algebras, except for the last section which contains a technical lemma (Lemma \ref{finiteNL}) whose proof essentially relies on Theorem \ref{takesakithm}.

\subsection{Hilbert Spaces and Operators}

We consider the reader familiar with the notions of Hilbert spaces and operators (continuous —or equivalently bounded— linear maps between Hilbert spaces). We refer to the classic textbooks of Conway \citeyear{conway} for the bases of the theory, and of Murphy \citeyear{murphy} for an excellent introduction to the theory of operator algebras. We will not dwell on the definitions and properties of von Neumann algebras, factors, and hyperfiniteness. We believe all these notions, though used in this paper and in Girard's, are not at the core of the characterization, and will not play an important rôle in the following construction. We therefore refer to the series of Takesaki \citeyear{takesaki1,takesaki2,takesaki3}. A quick overview of the needed material can also be found in the appendix of a paper by one the authors \cite{seiller-goim}.

We recall that an operator $T$ is a linear map from $\hil{H}$ —a Hilbert space— to $\hil{H}$ that is continuous. A standard result tells us that this is equivalent to $T$ being bounded, i.e. that there exists a constant $C$ such that for all $\xi\in\hil{H}$, $\norm{T\xi}\leqslant C\norm{\xi}$. The smallest such constant defines a norm on $\B{\hil{H}}$ —the set of operators on $\hil{H}$—which we will denote by $\norm{T}$.

Being given an operator $T$ in $\B{\hil{H}}$, we can show the existence of its \emph{adjoint} —denoted by $T^{\ast}$—, the operator that satisfies $\< T\xi,\eta\>=\<\xi,T^{\ast}\eta\>$ for all $\xi,\eta\in\hil{H}$. It is easily shown that $T^{\ast\ast}=T$, i.e. that $(\cdot)^{\ast}$ is an involution, and that it satisfies the following conditions:
\begin{enumerate}
\item For all $\lambda\in\mathbb{C}$ and $T\in\B{\hil{H}}$, $(\lambda T)^{\ast}=\bar{\lambda}T^{\ast}$;
\item For all $S,T\in\B{\hil{H}}$, $(S+T)^{\ast}=S^{\ast}+T^{\ast}$;
\item For all $S,T\in\B{\hil{H}}$, $(ST)^{\ast}=T^{\ast}S^{\ast}$.
\end{enumerate}

In a Hilbert space $\hil{H}$ there are two natural topologies, the topology induced by the norm on $\hil{H}$, and a weaker topology defined by the inner product.
\begin{enumerate}
\item The strong topology: we say a sequence $\{\xi_{i}\}_{i\in\mathbf{N}}$ converges strongly to $0$ when $\norm{\xi_{i}}\rightarrow 0$.
\item The weak topology: a sequence $\{\xi_{i}\}_{i\in\mathbf{N}}$ converges weakly to $0$ when $\<\xi_{i},\eta\>\rightarrow 0$ for all $\eta\in\B{\hil{H}}$. Weak convergence is thus a point-wise or direction-wise convergence.
\end{enumerate}
On $\B{\hil{H}}$, numerous topologies can be defined, each of which having its own advantages and drawbacks. The five most important topologies are the norm topology, the strong operator topology, the weak operator topology, the ultra-strong (or $\sigma$-strong) topology and the ultra-weak (or $\sigma$-weak) topology. We can easily characterize the first three topologies in terms of converging sequences as follows:
\begin{enumerate}
\item The norm topology: $\{T_{i}\}_{i\in\mathbf{N}}$ converges (for the norm) to $0$ when $\norm{T_{i}}\rightarrow 0$ ;
\item The strong operator topology, which is induced by the strong topology on $\hil{H}$: $\{T_{i}\}_{i\in\mathbf{N}}$ converges strongly to $0$ when, for any $\xi\in\hil{H}$, $T_{i}\xi$ converges strongly to $0$ ;
\item The weak operator topology, which is induced by the weak topology on $\hil{H}$: $\{T_{i}\}_{i\in\mathbf{N}}$ converges weakly to $0$ when, for any $\xi\in\hil{H}$, $T_{i}\xi$ converges weakly to $0$.
\end{enumerate}

We can show that $\B{\hil{H}}$ is the dual of a space denoted by $\B{\hil{H}}_{\ast}$ containing the \emph{trace-class operators}. For further details, the reader may refer to \cite{murphy} or \cite{takesaki1}. We remind here of this result only to define the $\sigma$-weak topology: if $A$ is a topological space and $A^{\ast}$ is its dual, the \emph{weak$^{\ast}$ topology} on $A$ is defined as the point-wise topology.

\subsection{Von Neumann Algebras in a Nutshell} 

Let $\hil{H}$ be a Hilbert space, and $\B{\hil{H}}$ be the set of bounded —continuous— linear maps from $\hil{H}$ to itself. It is standard knowledge that $\B{\hil{H}}$ is an associative algebra when endowed with composition and pointwise scalar multiplication and addition. It is moreover a complete normed vector space for the operator norm, defined as $\norm{u}=\sup\{x\in\hil{H}~|~\norm{u(x)}/\norm{x}\}$. It is therefore what is called a Banach algebra. On the other hand, it is known that every element of $\B{\hil{H}}$ has an \emph{adjoint} operator $u^{\ast}$. This operation $(\cdot)^{\ast}$ is an involution satisfying $(t+u)^{\ast}=t^{\ast}+u^{\ast}$, $(tu)^{\ast}=u^{\ast}t^{\ast}$, $(\lambda u)^{\ast}=\bar{\lambda}u^{\ast}$, $\norm{u^{\ast}}=\norm{u}$, and $\norm{u^{\ast}u}=\norm{u}^{2}$. A Banach algebra endowed with such an involution is called a C$^{\ast}$-algebra. As it turns out (this is the famous Gelfand-Naimark-Segal (GNS) construction), any C$^{\ast}$-algebra can be represented as a norm-closed $\ast$-subalgebra of $\B{\hil{H}}$ for a Hilbert space $\hil{H}$.

A von Neumann algebra $\vn{K}$ is a C$^{\ast}$-subalgebra of $\B{\hil{H}}$, where $\hil{H}$ is a Hilbert space, which is closed for a weaker topology than the norm topology: the strong-operator topology, which is pointwise convergence on $\hil{H}$ considered with its norm topology. The first important result of the theory, obtained by von Neumann, is that this requirement is equivalent to the requirement that $\vn{K}$ is closed for the even weaker \emph{weak operator topology} which is pointwise convergence on $\hil{H}$ considered with its weak topology. It is also equivalent to a completely algebraic condition which is the fact that $\vn{K}$ is equal to its bi-commutant: let us denote $\vn{K}'$ —the commutant of $\vn{K}$— the set of elements of $\B{\hil{H}}$ which commute with every element of $\vn{K}$, then $\vn{K}''$ denotes the bi-commutant of $\vn{K}$, that is the commutant of the commutant of $\vn{K}$.

The study of von Neumann algebras was quickly reduced to the study of factors, that is von Neumann algebras $\vn{K}$ whose center —the algebra of elements commuting with every element of $\vn{K}$— is trivial: i.e. von Neumann algebras $\vn{K}$ such that $\vn{K}\cap \vn{K}'=\complexN 1_{\vn{K}}$. Indeed, any von Neumann algebra can be decomposed along its center as a direct integral (a continuous direct sum) of factors. Factors $\vn{N}$ can then be easily classified by considering their sets of projections (operators $p$ such that $p=p^{\ast}=p^{2}$). Two projections $p,q$ are \emph{equivalent in $\vn{N}$} —denoted by $p\sim_{\vn{N}} q$— in the sense of Murray and von Neumann if there exists a partial isometry $u\in\vn{N}$ such that $uu^{\ast}=p$ and $u^{\ast}u=q$. A projection $p$ is \emph{infinite in $\vn{N}$} if there exists a proper subprojection $q<p$ (where $q\leqslant p$ is defined as $pq=q$, i.e. as the inclusion of the subspaces corresponding to $p$ and $q$) such that $p\sim_{\vn{N}} q$. A projection is \emph{finite} when it is not infinite. The classification of factor is as follows:
\begin{itemize}
\item \textbf{Type $\text{I}$:} $\vn{N}$ contains non-zero finite minimal projections. If the identity of $\vn{N}$ is the sum of a finite number —say $n$— of minimal projections, $\vn{N}$ is of type $\text{I}_{n}$, and if it is not the case $\vn{N}$ is of type $\text{I}_{\infty}$.
\item \textbf{Type $\text{II}$:}$\vn{N}$ contains finite projections but has no minimal projections. Then if the identity of $\vn{N}$ is a finite projection, $\vn{N}$ is of type $\text{II}_{1}$, and it is of type $\text{II}_{\infty}$ otherwise.
\item \textbf{Type $\text{III}$:}all the non-zero projections of $\vn{N}$ are infinite.
\end{itemize}

A typical example of type {I}$_{n}$ factor is the algebra of $n\times n$ matrices. Similarly, a typical example of type {I}$_{\infty}$ factor is the algebra $\B{\hil{H}}$ of bounded linear maps from a Hilbert space $\hil{H}$ to itself. Examples of type {II} and type {III} factors are more difficult to come by, and are generally constructed as von Neumann algebras defined from groups, or as von Neumann algebras induced by the (free and ergodic) action of a topological group acting on a measured space. Both these constructions are particular cases of the crossed product construction which is defined at the end of this section.

\begin{proposition}\label{matrixintofactor}
Any matrix algebra can be embedded in a type {II}$_{1}$ factor.
\end{proposition}

\begin{proof}
Let $k$ be an integer, and $\vn{M}$ denote the algebra of $k\times k$ matrices. Let $\vn{N}$ be a type {II}$_{1}$ factor. One can find in $\vn{N}$ a family $\pi_{1},\dots,\pi_{k}$ of projections such that $\sum_{i=1}^{k}\pi_{i}=1$ and which are equivalent in the sense of Murray and von Neumann, i.e. there exists partial isometries $(u_{i,j})_{1\leqslant i<j\leqslant k}$ such that $u_{i,j}u_{i,j}^{\ast}=\pi_{i}$ and $u_{i,j}^{\ast}u_{i,j}=\pi_{j}$. We will denote by $u_{j,i}$ the partial isometry $u_{i,j}^{\ast}$ and by $u_{i,i}$ the projection $\pi_{i}$. We can then define an embedding $\Psi$ of $\vn{M}$ into $\vn{N}$ as follows:
\begin{equation*}
(a_{i,j})_{1\leqslant i,j\leqslant k}\mapsto \sum_{i,j} a_{i,j}u_{i,j}
\end{equation*}
One can then easily check that $\Psi$ is a $\ast$-algebra injective morphism.
\end{proof}

Among von Neumann algebras, the approximately finite dimensional ones are of particular interest, and are usually called \emph{hyperfinite}. These are algebras in which every operator can be approximated (in the sense of the $\sigma$-weak topology\footnote{In a nutshell, the algebra $\B{\hil{H}}$ is the dual of the algebra of \emph{trace-class operators}. As a dual, it thus inherits the traditional weak$^{\ast}$ topology, which is called in the context of von Neumann algebras the $\sigma$-weak topology.}) by a sequence of finite-dimensional operators (elements of type $\text{I}_{n}$ factors, for $n\in\naturalN$). In particular, the type $\text{II}_{1}$ hyperfinite factor is unique up to isomorphism (in fact, hyperfinite factors of almost all types are unique).

The definition we gave of von Neumann algebras is a concrete definition, i.e. as an algebra of operators acting on a Hilbert space. It turns out that von Neumann algebras can be defined abstractly as C$^{\ast}$-algebras that are the dual space of a Banach space. In the next subsection, and more generally in this paper, the term \enquote{von Neumann algebra} will have the meaning of \enquote{abstract von Neumann algebra}.

\subsection{von Neumann Algebras and Groups}

\begin{definition}[Representations]
Let $\vn{K}$ be a von Neumann algebra. A couple $(\hil{H},\rho)$ where $\hil{H}$ is a Hilbert space and $\rho$ is a $\ast$-homomorphism from $\vn{K}$ to $\B{\hil{H}}$ is a \emph{representation} of $\vn{K}$. If $\rho$ is injective, we say the representation is \emph{faithful}.
\end{definition}

Among the numerous representations of a von Neumann algebra, one can prove the existence \cite{haagerup} of the so-called \emph{standard representation}, a representation satisfying several important properties.

The operation that will be of interest to us will be that of taking the \emph{crossed product} of an algebra and a group. This operation is closely related to that of semi-direct product of groups and is a way of internalizing automorphisms. Given an algebra $\vn{A}$ and a group $G$ of automorphisms of $\vn{A}$, we construct the algebra $\vn{A}\rtimes G$ generated by the elements of $\vn{A}$ and the elements of $G$.

\begin{definition}
An action of a topological group $G$ on a von Neumann algebra $\vn{K}$ is a continuous homomorphism of $G$ into $\text{Aut}(\vn{K})$.
\end{definition}

\begin{definition}[Crossed product (representations)]\label{crossedproduct}
Let $(\hil{H},\rho)$ be a representation of a von Neumann algebra $\vn{K}$, $G$ a locally compact group, and $\alpha$ an action of $G$ on $\vn{K}$. Let $\hil{K}=L^{2}(G,\hil{H})$ be the Hilbert space of square-summable $\hil{H}$-valued functions on $G$. We define representations $\pi_{\alpha}$ of $\vn{K}$ and $\lambda$ of $G$ on $\hil{K}$ as follows
\begin{equation*}
\begin{array}{rcll}
(\pi_{\alpha}(x).\xi)(g)&=&(\rho(\alpha(g)^{-1}(x))\xi(g)&(x\in\vn{K}, \xi\in\hil{K}, g\in G)\\
(\lambda(g).\xi)(h)&=&\xi(g^{-1}h)&(g,h\in G, \xi\in \hil{K})
\end{array}
\end{equation*}
Then the von Neumann algebra on $\hil{K}$ generated by $\pi_{\alpha}(\vn{K})$ and $\lambda(G)$ is called the crossed product of $(\hil{H},\rho)$ by $\alpha$.
\end{definition}
An important fact is that the result of the crossed product does not depend on the chosen representation of $\vn{K}$. The following theorem, which states this fact, will be of use in a technical lemma at the end of this paper.

\begin{theorem}[Unicity of the crossed product \protect{\cite[Theorem 1.7, p.~241]{takesaki2}}]\label{takesakithm}
Let $(\hil{H},\rho)$ and $(\hil{K},\rho')$ be two faithful representations of a von Neumann algebra $\vn{K}$, and let $G$ be a locally compact group together with an action $\alpha$ on $\vn{K}$. Then there exists an isomorphism between the crossed product of $(\hil{H},\rho)$ by $\alpha$ and the crossed product of $(\mathbb{K},\rho')$ by $\alpha$.

\end{theorem}

As a consequence, one can define \emph{the} crossed product of a von Neumann algebra and a group acting on it by choosing a particular representation. Of course, the natural choice is to consider the \emph{standard representation}.

\begin{definition}[Crossed product]
Let $\vn{K}$ be a von Neumann algebra, $G$ a group and $\alpha$ an action of $G$ on $\vn{K}$. The algebra $\vn{K}\rtimes_{\alpha} G$ is defined as the crossed product of the standard representation of $\vn{K}$ by $\alpha$.
\end{definition}

A particular case of crossed product is the crossed product of $\complexN$ by a (trivial) action of a group $G$. The resulting algebra is usually called the group von Neumann algebra $\N{G}$ of $G$. As it turns out, the operation of internalizing automorphisms of algebras (the crossed product) and the operation of internalizing automorphisms of groups (the semi-direct product) correspond: the algebra $\N{G\rtimes_{\alpha} H}$ is isomorphic to $\N{G}\rtimes_{\tilde{\alpha}} H$ where $\tilde{\alpha}$ is the action of $H$ on $\N{G}$ induced by the action of $H$ on $G$.

\section{Integers in the Hyperfinite Factor}\label{sec:integers}

\subsection{Binary Representation}\label{subsec:bin-rep}

We will embed the $(k+1)\times(k+1)$ matrices of Definition \ref{def_matricial_rep_integer} in the hyperfinite factor $\finhyp$ to have a uniform representation of the integers: an integer will be represented by an operator in $\vn{M}_{6}(\finhyp)$ fulfilling some properties. To express them we define, given a sequence $\star a_{1} \dots a_{k}$ representing an integer $n$ and for $j, l \in \{0, 1\}$, the sets:
\begin{eqnarray*}
I_{jl}^{n}&=&\{1\leqslant i\leqslant k ~|~ a_{i}= j, a_{i+1}=l\}\\
I_{Sj}^{n}&=&\{i=1 ~|~ a_{i}=j\}\\
I_{jE}^{n}&=&\{i=k ~|~ a_{i}=j \}
\end{eqnarray*}
Roughly speaking, $I_{Sj}^n$ (resp. $I_{jE}^n$) tells us about the first (resp. last) bit of $n$ and $I_{jl}^n$ is the set of sequences of a $j$ followed by a $l$.

\begin{definition}[Binary representation of integers]\label{binrep}
An operator $N_n\in\vn{M}_{6}(\finhyp)$ is a \emph{binary representation} of an integer $n$ if there exists projections $\pi_{0},\pi_{1},\dots,\pi_{k}$ in $\finhyp$ that satisfy $\sum_{i=0}^{k}\pi_{i}=1$ such that:
\begin{equation*}
N_{n}=\left(\begin{array}{cccccc}
	0 & l_{00} & 0 & l_{10} & l_{S0} & 0\\
	l_{00}^{\ast} & 0 & l_{01}^{\ast} & 0 & 0 & l_{0E}^{\ast}\\
	0 & l_{01} & 0 & l_{11} & l_{S1} & 0\\
	l_{10}^{\ast} & 0 & l_{11}^{\ast} & 0 & 0 & l_{1E}^{\ast}\\
	l_{S0}^{\ast} & 0 & l_{S1}^{\ast} & 0 & 0 & 0\\
	0 & l_{0E} & 0 & l_{1E} & 0 & 0
	\end{array}\right)	
\end{equation*}
where the coefficients are partial isometries fulfilling the equations (where $\pi_{k+1}=\pi_{0}$):
\begin{eqnarray*}
l_{\star}&=&\sum_{i\in I_{\star}^{n}} \pi_{i+1}l_{\star}\pi_{i}~~~~(\star\in\{00,01,10,11,S0,S1,0E,1E\})\\
\pi_{0}&=&(l_{0E}+l_{1E})(l_{00}+l_{01}+l_{10}+l_{11})^{k-1}(l_{S0}+l_{S1})
\end{eqnarray*}
\end{definition}
\begin{proposition}[Binary and matricial representations]\label{matricerepbinaire}
Given $N_n\in\vn{M}_{6}(\finhyp)$ a binary representation of the integer $n$, there exists an embedding $\theta:\vn{M}_{k+1}(\complexN)\rightarrow\finhyp$ such that\footnote{We denote by $\text{Id}$ the identity matrix of $\vn{M}_{6}(\complexN)$. We will allow ourselves the same abuse of notation in the following statements and proofs in order to simplify the formulas.} $\text{Id}\otimes\theta(M_n)=N_n$, where $M_n$ is the matricial representation of $n$.
\end{proposition}
\begin{proof}
Let $N_n\in\finhyp$ a binary representation of $n\in\naturalN$, and $\pi_{0},\dots,\pi_{k}$ the associated projections. Notice that the projections $\pi_{i}$ are pairwise equivalent.

We now define an embedding $\theta : \vn{M}_{k+1}(\complexN) \rightarrow \finhyp$:
\begin{equation*}
\theta: (a_{i,j})_{0\leqslant i,j\leqslant k}\mapsto \sum_{i=0}^{k}\sum_{j=0}^{k} a_{i,j}u_{i,j}
\end{equation*}
with:
\begin{equation*}
u_{i,j}=\left\{\begin{array}{ll}
		(l_{00}+l_{01}+l_{10}+l_{11})^{j-1}(l_{S0}+l_{S1})&\text{ if }i=0\\
		(l_{00}+l_{01}+l_{10}+l_{11})^{j-1}&\text{ if $i<j$ and }i\neq0\\
		((l_{00}+l_{01}+l_{10}+l_{11})^{i-1}(l_{S0}+l_{S1}))^{\ast}&\text{ if } j=0\\
		((l_{00}+l_{01}+l_{10}+l_{11})^{i-1})^{\ast}&\text{ if $i>j$ and }j\neq0\\
		\pi_{k}&\text{ if }i=j=k
		\end{array}\right.		
\end{equation*}
We can easily check that the image by $\text{Id}\otimes\theta$ of the matrix $M_n$ representing $n$ is equal to $N_n$.
\end{proof}

This new representation is a gain in terms of uniformity, as all the integers are represented by matrices of the same size. But at the same time, as any embedding $\theta :\vn{M}_{k+1}(\complexN) \rightarrow \finhyp$ define a representation of the integers, we have to check that they all are equivalent (Proposition \ref{entiersbinairesunitaire}) and to define (Definition \ref{defnormpair}) a framework where the representation of the integers and the programs can interact as expected.

\begin{proposition}[Equivalence of binary representations] \label{entiersbinairesunitaire}
Given $N_n$ and $N_n'$ two binary representations of $n\in \naturalN$, there exists a unitary $u\in\finhyp$ such that $(\text{Id}\otimes u)N_n(\text{Id}\otimes u)^{\ast}=N_n'$.
\end{proposition}

\begin{proof}
Let $\pi_{0},\dots,\pi_{n}$ (resp. $\nu_{0},\dots,\nu_{n}$) be the projections and $l_{\star}$ (resp. $l'_{\star}$) the partial isometries associated to $N_n$ (resp. $N_n'$). It is straightforward that $\pi_{0}$ and $\nu_{0}$ are equivalent according to Murray and von Neumann definition, so there exists a partial isometry $v$ such that $vv^{\ast}=\nu_{0}$ and $v^{\ast}v=\pi_{0}$. For all $ 0 \leqslant i \leqslant n$ we define the partial isometries:
\begin{equation*}
v_{i}=((l'_{00}+l'_{01}+l'_{10}+l'_{11})^{i-1}(l'_{S0}+l'_{S1}))v((l_{00}+l_{01}+l_{10}+l_{11})^{i-1}(l_{S0}+l_{S1}))^{\ast}
\end{equation*}
We can easily check that:
\begin{eqnarray*}
v_{i}v_{i}^{\ast}&=&\nu_{i}\\
v_{i}^{\ast}v_{i}&=&\pi_{i}
\end{eqnarray*}
It follows that the sum $u=\sum_{i=0}^{n} v_{i}$ is a unitary and $(\text{Id}\otimes u)N_n(\text{Id}\otimes u)^{\ast}=N_n'$.
\end{proof}

\subsection{Normative Pairs}\label{subsec:norm_pair}
The notion of \emph{normative pair}, a pair of two subalgebras $(\vn{N},\vn{O})$, was defined by Girard \cite{normativity} in order to describe the situations in which an operator in $\vn{O}$ acts uniformly on the set of all representations of a given integer in $\vn{N}$. Indeed, as we just explained, we no longer have uniqueness of the representation of integers. An operator representing a kind of abstract machine should therefore interact in the same way with different representations of the same integer. 

The notion of normative pair therefore depends on the notion of interaction one is considering. The interaction used by Girard was based on Fuglede-Kadison determinant\footnote{A generalization of the usual determinant of matrices that can be defined in a type $\text{II}_{1}$ factor.}. As a matter of fact, Girard defines his interaction with the determinant but actually uses nilpotency in his proofs. In order to give more flexibility to the definitions, we chose to work with an interaction based on nilpotency, which represents the fact the computation ends. This change in definition does not modify the fact that one can characterize \cc{co-NL}, but allows one to consider a broader class of groups\footnote{The use of the determinant forces Girard to consider only amenable groups, so that the result of the crossed product in Proposition \ref{croisenormative2} yields the type $\text{II}_{1}$ hyperfinite factor.}, and a broader class of languages\footnote{In this paper and in Girard's, we consider languages obtained from finite positive linear combinations of unitaries induced by the group elements. The positivity of the coefficients is needed so that the condition involving the determinant implies the nilpotency. However, these conditions are no longer equivalent if one allows negative coeficients. As a consequence, this new definition of normative pair extends the number of languages that can be defined.}.



\begin{definition}[Normative Pairs]\label{defnormpair}
Let $\vn{N}$ and $\vn{O}$ be two subalgebras of a von Neumann algebra $\vn{K}$. The pair $(\vn{N},\vn{O})$ is a \emph{normative pair (in $\vn{K}$)} if:
\begin{itemize}
\item $\vn{N}$ is isomorphic to $\finhyp$;
\item For all $\Phi\in\vn{M}_{6}(\vn{O})$ and $N_n,N_n'\in\vn{M}_{6}(\vn{N})$ two binary representations of $n$,
\begin{equation*}
\Phi N_n\text{ is nilpotent}\Leftrightarrow \Phi N_n'\text{ is nilpotent}
\end{equation*}
\end{itemize}
\end{definition}


\begin{proposition}\label{croisenormative2}
Let $S$ be a set and for all $s\in S$, $\vn{N}_{s}=\finhyp$. For all group $G$ and all action $\alpha$ of $G$ on $S$, the algebra $\vn{K}=(\bigotimes_{s\in S} \vn{N}_{s})\rtimes_{\hat{\alpha}} G$ -- where $\hat{\alpha}$ denotes the action induced by $\alpha$ on the tensor product -- contains a subalgebra generated by $G$ that we will denote $\vn{G}$. Then for all $s\in S$, the pair $(\vn{N}_{s},\vn{G})$ is a normative pair (in $\vn{K}$).
\end{proposition}


\begin{proof}
From the hypotheses, $\vn{N}_{s}$ is isomorphic to $\finhyp$. Regarding the second condition, we will only show one implication, the other being obtained by symmetry. By Lemma \ref{entiersbinairesunitaire}, there exists a unitary $u$ such that $(\text{Id}\otimes u)N_n(\text{Id}\otimes u)^{\ast}=N_n'$. We define $v=\bigotimes_{s\in S} u$ and $\pi_{v}$ the unitary in $\vn{K}$ induced by $v$. Then $\pi_{v}$ commutes with the elements of $\vn{G}$, so if there exists $d \in \naturalN$ such that $(\phi N_n)^{d}=0$, then $ (\phi N_n')^{d}=(\phi uN_nu^{\ast})^{d}=(u\phi N_nu^{\ast})^{d}=u(\phi N_n)^{d}u^{\ast}=0$. 
\end{proof}

\begin{definition}[Observations]
Let $(\vn{N},\vn{G})$ be a normative pair. An \emph{observation} is an operator in $\vn{M}_6 (\vn{G}) \otimes \vn{Q}$, where $\vn{Q}$ is a matrix algebra, i.e. $\vn{Q}=\vn{M}_{s}(\complexN)$ for an integer $s$, called the \emph{algebra of states}.
\end{definition}

\begin{definition}
Let $(\vn{N},\vn{G})$ be a normative pair, and $\phi$ an observation. We define the set of natural numbers:
\begin{equation*}
[\phi]=\{n\in\naturalN~|~\phi N_{n}\text{ is nilpotent, $N_{n}$ a binary representation of $n$}\}
\end{equation*}
\end{definition}

\begin{definition}
Let$(\vn{N}_{0},\vn{G})$ be a normative pair and $X\subset \cup_{i=1}^{\infty} \vn{M}_{6}(\vn{G})\otimes\vn{M}_{i}(\complexN)$ be a set of observations. We define the \emph{language decided by $X$} as the set:
\begin{equation*}
\{X\}=\{[\phi]~|~\phi\in X\}
\end{equation*}
\end{definition}

\begin{corollary}\label{pairenormG}
Let $\mathfrak{S}$ be the group of finite permutations over $\naturalN$, and for all $n\in\mathbf{N}$, $\vn{N}_{n}=\finhyp$. Then $(\vn{N}_{0},\vn{G})$ is a normative pair in $(\bigotimes_{n\in \naturalN} \vn{N}_{n})\rtimes_{\hat{\alpha}} \mathfrak{S}$.
\end{corollary}

In this particular case, the algebra $(\bigotimes_{n\in\naturalN} \vn{N}_{n})\rtimes_{\hat{\alpha}} \mathfrak{S}$ is the type $\text{II}_{1}$ hyperfinite factor. This is one of the reason why Girard considered it, as it is then possible to use Fuglede-Kadison determinant. From now on, we will consider this normative pair fixed, and we will study two sets of observations.

\begin{definition}[$P_{\geqslant 0}$ and $P_{+}$]
An observation $(\phi_{i,j})_{0\leqslant i,j\leqslant 6s}\in\vn{M}_{6}(\vn{G})\otimes\vn{M}_{s}(\complexN)$ is said to be \emph{positive} (resp. \emph{boolean}) when for all $i,j$, $\phi_{i,j}$ is a positive finite linear combination (resp. a finite sum) of unitaries induced by elements of $\mathfrak{S}$, i.e. $\phi_{i,j}=\sum_{k\in I_{i,j}} \alpha_{i,j}^{k}\lambda(\sigma_{i,j}^{k})$ with $\alpha_{i,j}^{k}\geqslant 0$ (resp. $\alpha_{i,j}^{k}=1$).

We then define the following sets of observations:
\begin{eqnarray*}
P_{\geqslant 0}&=&\{\phi~|~\phi\text{ is a positive observation}\}\\
P_{+}&=&\{\phi~|~\phi\text{ is a boolean observation}\}
\end{eqnarray*}
\end{definition}

It is not clear at this point how a program could be expressed as an observation. In the next section, we will introduce a notion of abstract machines which is well suited to be represented as an observation. We will then show how one can define an observation that simulates such a machine.

\section{Non-Deterministic Pointer Machines}\label{sec:ndpm}

We define in this section the notion of non-deterministic pointer machines (\NDPM), an abstract device really close to the multi-head finite automata \cite{rosenberg66}, well known to characterize logspace computation.
The two have in common the fact that they may only move a fixed number of pointers, read the pointed values and according to their (non-deterministic) transition function change the position of the pointers and their state.

However, we felt it necessary to introduce this model of computation because it has several peculiarities that will help encode them as operators:
\begin{itemize}
\item It is \enquote{universally non-deterministic}: if one branch of computation rejects, the whole computation rejects. It is convenient because acceptance is represented as the nilpotency of an operator.
\item Acceptance and rejection are in the codomain of the transition function, and not states, because we want the computation to stop or to loop immediately, and not to have to define the \enquote{last movement} of the pointers.
\item The alphabet is fixed to $\{0, 1, \star\}$, because these are the only values encoded in the binary representation of the integers.
\item Its input is circular, because in the binary representation we can access both the last and first bits of the integer from the symbol $\star$.
\item The \enquote{initial configuration} (in fact, the pseudo-configuration, defined below) is a parameter that will be used to make the operator loop properly.
\item The values are read and stored only when the pointer move, because before the computation starts, the operator cannot access the input.
\item If the transition relation is not defined for the current situation, the \NDPM accepts, because that's the way the operator will behave.
So acceptation is the \enquote{default} behavior, whereas rejection is meaningfull.
We could equivalently have forced the relation transition to be total.
\end{itemize}
Moreover, we will prove in the following that \NDPMs can be modified to always halt, and to move at most one pointer at a time.

This device may remind of the programming language PURPLE \cite{pointer08} as we cannot remember any value nor access the address of the pointers, and it may be interesting to study the relations between the latter and our machines. However, since this paper is focused on the study of a non-deterministic framework\footnote{Since the writing of this paper, an article dealing with a non-deterministic variant of PURPLE has been published \cite{Hofmann2013}.}, we postpone this question to a future work dealing with deterministic complexity classes. Here, we will focus on proving that \NDPMs can recognize any \cc{co-NL} set.

A pointer machine is given by a set of pointers that can move back and forth on the input tape and read (but not write) the values it contains, together with a set of states. For $1 \leqslant i \leqslant p$, given a pointer $p_{i}$, only one of three different \emph{instructions} can be performed at each step:
\begin{center}
$p_{i}+$, i.e. \enquote{move one step forward},\\
$p_{i}-$, i.e. \enquote{move one step backward},\\
$\epsilon_{i}$, i.e. \enquote{do not move}
\end{center}
We define the set of instructions $I_{\{1,\dots,p\}}=\{p_{i}+, p_{i}-, \epsilon_{i}~|~i\in\{1,\dots,p\}\}$. We will denote by $\pos{p_i}$ the value of the pointer (the address it points at), that is the number of cells clockwise between $\star$ and the bit pointed by $p_i$, i.e. the \emph{distance} between $\star$ and the bit pointed. Note that the alphabet $\Sigma$ is fixed to $\{ 0, 1, \star\}$.

\begin{definition}
A non-deterministic pointer machine with $p \in \naturalN^*$ pointers is a pair $M=(Q, \rightarrow)$ where $Q$ is the set of \emph{states} and $\rightarrow \subseteq (\{ 0, 1, \star\}^p \times Q) \times ((I_{\{1,\dots,p\}}^p \times Q) \cup \{\textbf{accept, reject}\})$ is the \emph{transition relation}. We write \NDPMp{p} the set of \NDPMs with $p$ pointers.
\end{definition}

We define a \emph{pseudo-configuration} $c$ of $M \in \NDPMp{p}$ as a \enquote{partial snapshot}: an element in $\{0,1,\star\}^p \times Q$ that contains the last values read by the $p$ pointers and the current state, \emph{but does not contain the addresses of the $p$ pointers}. The set of pseudo-configurations of a machine $M$ is written $C_M$ and it is the domain of the transition relation. 

Let $M \in \NDPMp{p}$, $c \in C_M$ and $n\in \naturalN$ an input. We define $M_{c}(n)$ as $M$ with $n$ encoded as a string on its circular input tape (as $\star a_1 \hdots a_k$ for $a_1 \hdots a_k$ the binary encoding of $n$ and $a_{k+1}=a_0=\star$) starting in the pseudo-configuration $c$ with $\pos{p_{i}}=0$ for all $1 \leqslant i \leqslant p$ (that is, the pointers are initialized with the address of the symbol $\star$). The pointers may be considered as variables that have been declared but not initialized yet. They are associated with \emph{memory slots} that store the values and are updated only when the pointer moves, so as the pointers did not moved yet, those memory slots haven't been initialized. The \emph{initial pseudo-configuration} $c$ initializes those $p$ registers, not necessarily in a faithful way (it may not reflect the values contained at $\pos{p_i}$). The entry $n$ is \emph{accepted} (resp. \emph{rejected}) by $M$ with \emph{initial pseudo-configuration} $c \in C_M$ if after a finite number of transitions every branch of $M_c(n)$ reaches \textbf{accept} (resp. at least a branch of $M$ reaches \textbf{reject}). We say that $M_c(n)$ halts if it accepts or rejects $n$ and that $M$ decides a set $S$ if there exists an initial pseudo-configuration $c \in C_M$ such that $M_c(n)$ accepts if and only if $n \in S$. We write $\lang{M}$ the set decided by $M$.

\begin{definition}
Let \cc{\{NDPM\}} be the class of sets $S$ such that there exists a \NDPM that decides $S$.
\end{definition}

\paragraph{\textbf{One movement at a time}}We can prove that for all $M \in \NDPMp{p}$ there exists $M \in \NDPMp{p}$ such that for all $\sigma_1, \hdots, \sigma_p, \textbf{q} \rightarrow' p_1, \hdots, p_p, \textbf{q}'$ at most one instruction among $p_1, \hdots, p_p$ differs from $\epsilon_i$ —stated informally, such that no more than one pointer moves at every transition— and such that $\lang{M}=\lang{M'}$. 
The number of states of $M'$ and the number of transitions needed by $M'$ to decide the same set increase, but that does not affect our machine in terms of complexity as the number of transitions and the cardinality of $Q$ will not be measures of the complexity of our machines.


\paragraph{\textbf{Shorthands}} We use the symbol $*$ for any symbol among $\{0, 1, \star\}$, $0/1$ for \enquote{$0$ or $1$}. For instance $(*, 0, \textbf{q}) \rightarrow (\epsilon_1, p_2+, \textbf{q}')$ will be a shorthand for
\begin{eqnarray*}
(0, 0, \textbf{q}) &\rightarrow &(\epsilon_1, p_2+, \textbf{q}')\\
(1, 0, \textbf{q}) &\rightarrow &(\epsilon_1, p_2+, \textbf{q}')\\
(\star, 0, \textbf{q}) &\rightarrow &(\epsilon_1, p_2+, \textbf{q}')
\end{eqnarray*}

\paragraph{\textbf{\emph{Sensing pointers}}}
We can easily mimic \enquote{sensing pointers}, i.e. answer the question \enquote{\emph{Is $\pos{p_1}=\pos{p_2}$?}}, by the help of the following routine, which need a third pointer $p_3$ with $\pos{p_3}=0$. At every transition, $p_1$ and $p_2$ move one square left and $p_3$ moves one square right. Two cases arise:
\begin{itemize}
\item $p_1$ and $p_2$ reach $\star$ after the same transition,
\item $p_1$ (or $p_2$) reaches $\star$ whereas the other pointer is not reading $\star$.
\end{itemize}
According to the situation, we encode that they were at the same position or not in the state. Then $p_1$ and $p_2$ moves at each transition one square right, $p_3$ moves at each transition one square left, and when $\pos{p_3}=0$, we resume the computation. We can easily check that $p_1$ and $p_2$ are back to their initial position, and now we can retrieve from the state if they were at the same position or not, i.e. if we had $\pos{p_1}=\pos{p_2}$. Notice moreover that $p_3$ is back to $\star$ and ready to be used for another comparison.

\paragraph{\textbf{To express any number}}
It is possible to express a distance superior to the size $k$ of the input to a routine: $j$ pointers can represent a distance up to $k^j$. Every time the $i$-th pointer made a round-trip (that is, is back on $\star$), the $i+1$-th pointer goes one cell right. By acting like the hands of a clock, the $j$ pointers can encode any integer inferior to $k^j$.

To decode the distance expressed by $j$ pointers $p_1, \hdots, p_j$, it is sufficient to have $j$ pointers $p'_1, \hdots, p'_j$ and to move them clockwise until for all $1 \leqslant i \leqslant j$, $\pos{p_i}=\pos{p'_i}$.

We will for the sake of simplicity consider that any distance $\bigO(k^j)$ can be expressed by a single pointer, even if it may require several pointers to be properly expressed. We make this idea formal in the proof of Lemma \ref{lelemmedetransfo}, by defining how to implement a clock in any \NDPM.

\paragraph{\textbf{Pointer arithmetic}}
It is classical pointer arithmetic to prove that with the help of some additional pointers, \NDPMs can compute addition, subtractions, multiplication, division, logarithm and modulo, i.e. that given two pointers $p_1$ and $p_2$, it is possible to let a third pointer $p_3$ be at $\pos{p_1} + \pos{p_2}$, $\pos{p_1} - \pos{p_2}$, $\pos{p_1} \times \pos{p_2}$, $\lfloor \pos{p_1} / \pos{p_2} \rfloor$, $\lceil \log(\pos{p_1})\rceil$ or $\pos{p_1}\bmod{\pos{p_2}}$.
Needless to say, those operations permit to establish bit by bit the binary expression of an integer encoded by $\pos{p}$.

We will only deal with decision problems and ask ourselves what sets can be recognized in this framework.
It turns out that we can recognize any \cc{co-NL}-set, and to prove it we will use the most common method\footnote{That can be reminded to the reader in \cite{arora2009}, pp. 88--89.}: we will exhibit a \NDPM that can solve a \cc{co-NL}-complete problem, and define another \NDPM that reduce any \cc{co-NL} problem to this \cc{co-NL}-complete problem.

\begin{definition}[\stconncomp]
We define the following problem: \enquote{given a (directed) graph encoded as a list of adjacences, accept if and only if there is \textbf{no} path from the source (numbered $1$) to the target (numbered $n$) in the graph}. This problem, known as \stconncomp or \pb{REACHABILITYComp}, is \cc{co-NL} complete. We define the set 
\[\stconncomp= \{n \in \naturalN~|~ n \text{ does \textbf{not} encode a graph where there is a path from }1 \text{ to } n \}\]
\end{definition}

\begin{proposition} \label{M_decide}
$\stconncomp \in \cc{\{NDPM\}}$
\end{proposition}

\begin{proof}\label{proof_stconn}
Given a graph of size $n$, the input will be encoded as
\[\star\underbrace{00\hdots00}_{n \text{ bits}}\fbox{1}\overbrace{a_{11}0a_{12}0\hdots 0a_{1n-1}0a_{1n}}^{\text{edges going from}1}\fbox{1}\dots\fbox{1}\overbrace{a_{n1}0a_{n2}0\hdots 0 a_{nn-1}0a_{nn}}^{\text{edges going from }n}\fbox{1}\]
where $(a_{ij})$ is the adjacency list, that is to say that $a_{ij}=1$ if there is an edge from the vertex numbered by $i$ to the vertex numbered by $j$, $0$ elsewhere. The boxed bits in the figure above are \enquote{separating} bits, between the coding of $n$ and the list of adjacences, and between the coding of the edges of source $i$ and the coding of the edges of source $i+1$.

We define a \NDPM $M$ such that $M_c(n)$ with $c=\{\star, \star, \star, \star, \textbf{Init}\}$ accepts if and only if $n \in$ \stconncomp.

The transition relation of $M$ is presented in the figure \ref{fig:transition_fction}. Informally, our algorithm goes as follow:

The pointer $p_1$ counts the size of the path followed. Every time we follow an edge, we move $p_1$ forward on the string made of $n$ bits (second line of \ref{nondet}). The pointer $p_2$ will scan the encoding of the outgoing edges of a vertex, \enquote{followed} by $p_3$: when $p_2$ is reading $a_{ij}$ then $p_3$ will be at $a_{j1}$. If $a_{ij}=1$ (premise of \ref{nondet}), a non-deterministic transition takes place: on one hand we continue to scan the outgoing edges from $i$, on the other we increment $p_1$, place $p_2$ at $a_{j1}$ and $p_3$ at $a_{11}$.
The pointer $p_4$ \enquote{follows} $p_3$ on the $n$ first bits, and if $p_4$ reaches a $1$ when $p_2$ reads that there is an edge, it means that there is an edge whose target is $n$, and so we reject (\ref{reject1}). When $p_2$ finishes to browse the adjacency list of an edge, we accept (\ref{accept1}). If $p_1$ reaches a $1$ and $p_4$ reads a $0$ (premise of \ref{accept2}), it means that we already followed $n$ edges without ever targeting the vertex $n$, so we end up accepting. As we know that if there is a path from $1$ to $n$ then there exists a path of size at most $n$, $M_c(n)$ will accept if and only if $n \in \stconncomp$, elsewhere $M_c(n)$ rejects.

\begin{figure}
\centering
\begin{align}
(\star, \star, \star, \star, \textbf{Init}) &\rightarrow (p_1+, p_2+, p_3+, p_4+, \textbf{Init})\\
(*, 0, *, *, \textbf{Init}) &\rightarrow (\epsilon_1, p_2+, p_3+,\epsilon_4, \textbf{Init}) \\
(*, 1, *, *,\textbf{Init}) &\rightarrow (\epsilon_1, p_2+, \epsilon_3,\epsilon_4, \textbf{out.edge?})\\ 
(*, 0, *, *,\textbf{out.edge?}) &\rightarrow (\epsilon_1, p_2+, \epsilon_3,p_4+, \textbf{no.edge}) \label{out.edge?}\\
(*, 0, *, *,\textbf{no.edge}) &\rightarrow (\epsilon_1, \epsilon_2, p_3+,\epsilon_4, \textbf{p3.next.node})\\
(*, 1, *, *,\textbf{no.edge}) &\rightarrow \textbf{accept} \label{accept1}\\
(*, *, *, *,\textbf{p3.next.node}) &\rightarrow (\epsilon_1, \epsilon_2, p_3+, \epsilon_4,\textbf{reading.sep.bit})\\
(*, *, 0, *,\textbf{reading.sep.bit}) &\rightarrow (\epsilon_1, \epsilon_, p_3+,\epsilon_4, \textbf{p3.next.node}) \label{sep.bit1} \\
(*, *, 1, *,\textbf{reading.sep.bit}) &\rightarrow (\epsilon_1, p_2+, \epsilon_3, \epsilon_4, \textbf{out.edge?}) \label{sep.bit2} \\
(*, 1, *, *,\textbf{out.edge?}) &\rightarrow \begin{cases}(\epsilon_1, p_2+, \epsilon_3, p_4+, \textbf{no.edge})\\
(p_1+, \epsilon_2, \epsilon_3, p_4+, \textbf{edge.found}) \end{cases}\label{nondet}\\
(*, *, *, 1,\textbf{edge.found}) & \rightarrow \textbf{reject} \label{reject1}\\
(1, *, *, 0,\textbf{edge.found}) & \rightarrow \textbf{accept} \label{accept2}\\
(*, *, *, 0,\textbf{edge.found}) & \rightarrow (\epsilon_1, p_2-, \epsilon_3, p_4-, \textbf{rewind.p2.p4})\\
(*, *, *, 0/1,\textbf{rewind.p2.p4}) & \rightarrow (\epsilon_1, p_2-, \epsilon_3, p_4-, \textbf{rewind.p2.p4}) \label{noterewindp2}\\
(*, *, *, \star,\textbf{rewind.p2.p4}) & \rightarrow (\epsilon_1, p_2-, \epsilon_3, \epsilon_4, \textbf{rewind.p2})\\
(*, 0/1, *, *,\textbf{rewind.p2}) & \rightarrow (\epsilon_1, p_2-, \epsilon_3, \epsilon_4, \textbf{rewind.p2})\\
(*, \star, *, *,\textbf{rewind.p2}) & \rightarrow (\epsilon_1, p_2+, p_3-, \epsilon_4, \textbf{exchange.p2.p3.})\\
(*, *, 0/1, *,\textbf{exchange.p2.p3}) & \rightarrow (\epsilon_1, p_2+, p_3-, \epsilon_4, \textbf{exchange.p2.p3.})\\
(*, *, \star, *,\textbf{exchange.p2.p3}) & \rightarrow (\epsilon_1, \epsilon_2, p_3+, \epsilon_4, \textbf{get.p3.to.start})\\
(*, *, 0, *,\textbf{get.p3.to.start}) & \rightarrow (\epsilon_1, \epsilon_2, p_3+, \epsilon_4, \textbf{get.p3.to.start}) \\
(*, *, 1, *,\textbf{get.p3.to.start}) & \rightarrow (\epsilon_1, p_2+, \epsilon_3, \epsilon_4, \textbf{out.edge?})
\end{align}
\caption{The transition relation to decide \stconncomp}
\label{fig:transition_fction}
\end{figure}
\end{proof}

We now adapt the classical logspace-reduction from any problem in \cc{NL} to \stconn. Given a \cc{co-NL}-problem \pb{Pb}, there exists a non-deterministic logspace Turing Machine $M$ that decides it. To solve \pb{Pb} is just to establish if there is \textbf{no} transition from the initial configuration to a rejecting configuration of $M$, once the computational graph of $M$ is given. We now make some assumptions on $M$ and prove how a \NDPM can output any bit of the transition graph of $M$.

Given any set $\pb{Pb} \in \cc{co-NL}$, there exists a non-deterministic logspace Turing Machine $M$ such that $M$ accepts $n$\footnote{Meaning that \textbf{all} branches reach \textbf{accept} after a finite number of transitions.} iff $n \in \pb{Pb}$. We can assume w.l.o.g. that $M$ works on the alphabet $\Sigma=\{0, 1\}$, does not cycle, always halts, has one read-only tape and one read-write working tape whose precise bound is $k \times (\log(\sizeof{n}))$.
Those are classical \enquote{hacking} of Turing Machines that should not surprise the reader.
We may also assume that the names of the states are written in binary, so that for $\sizeof{Q}=q$ the number of states of $M$, any state may be written with $\lceil \log{q} \rceil$ bits. At last, we may assume that the instructions to move the heads are written with two bits. All those assumptions make clear that $M$ may be entirely described with a binary string. 

We know that $M(n)$ has less than
\[2^{(k \times (\log(\sizeof{n})))} \times (k \times (\log(\sizeof{n}))) \times (\log(\sizeof{n})) \times \lceil \log(q) \rceil\]
different configurations.
It reflects respectively the content of the working tape, the position of the read-write and read-only heads and the state.
This is equivalent to $2^{\mathcal{O}(\log(|n|))}$, so we know there exists a $d$ such that $M(n)$ has less than $\sizeof{n}^d$ different configurations.

Any configuration of $M(n)$ may be described as
\[
\underbrace{01000\hdots010\hdots011}_{\textnormal{Position of the read head}}\overbrace{\sigma 0\sigma 0 \hdots \sigma 0 \sigma 1 \sigma 0 \hdots \sigma 0 \sigma 0 \sigma 0 \sigma 0 \sigma 0 \sigma 0 \sigma 0}^{\text{Working tape and position of the working head}}\underbrace{01\hdots10}_{\text{state}}
\]
where the $\lceil \log(\sizeof{n}) \rceil$ first bits encode the position of the reading head in binary, $\sigma$ corresponds to the bits on the working tape and the bit that follows $\sigma$ equals $1$ iff the working head is on that cell.
The remaining $\lceil \log(q) \rceil$ bits express the current state.

This binary string is of length $\lceil \log(\sizeof{n}) \rceil \times (2 \times (\lceil \log{\sizeof{n}} \rceil \times k)) \times \lceil \log(q) \rceil $, i.e. there exists a $e$ such that this string is of length inferior to $e \times \log(\sizeof{n})^2$.

Among all the binary strings of size $e \times \log(\sizeof{n})^2$, some correspond to configurations, and some do not (for instance because the working head is supposed to be in several places at the same time) -- we will call them \enquote{phantom configurations}.

The \emph{configuration graph} of $M$ on input $n$ is simply the graph where configurations are vertices, and there is an edge between two vertices iff there is a transition in $M(n)$ between the two corresponding configurations.

\begin{lemma}[Pointer-reduction]
\label{point_lemma}
For all non-deterministic logspace Turing Machine $M$, there exists a \NDPM $T$ such that for all $n$, given a pointer $p_d$ with $\pos{p_d}=j$, $T$ accepts iff the $j$-th bit of the encoding of the computation graph of $M$ on input $n$ is $1$, rejects if it is $0$.
\end{lemma}

\begin{proof}
Recall we use the encoding of the proof of Proposition \ref{M_decide} to express the encoding of the configuration graph of $M$.
The \NDPM $T$ will act as a \enquote{transducer} as follows:

\begin{itemize}
\item It computes the number of binary strings of size $e \times \log(\sizeof{n})^2$.
This number is bounded by $2^{e \times \log(\sizeof{n})^2}$ and we saw previously that a \NDPM could express such distances.
Then $T$ compares this value to $j$ : if $j$ is inferior, it rejects, if $j$ is equal, it accepts, elsewhere it goes on.
This reflects the initial bits set to $0$ to express in unary the size of the graph.
\item Elsewhere $T$ establishes if $j$ corresponds to a \enquote{separating bit} and accepts or rejects accordingly, that can be simply made with the division and modulo operations.
\item Elsewhere, $j$ encodes a query regarding the presence or absence of transition between two configurations $a$ and $b$.
If $a=b$, there is no need to explore this transition\footnote{Because that would imply that there is a transition from a configuration to itself, and so $M(n)$ is stuck in a loop.}, and $T$ rejects.
Elsewhere $T$ establishes if there is a transition between $a$ and $b$, and accepts or rejects accordingly.
\end{itemize}
This last point needs to be made more precise: if $j > 2^{e \times \log(\sizeof{n})^2}$ and if $j$ does not correspond to a \enquote{separating bit}, it means that the value of $j$ corresponds to the absence or presence of an edge between two vertices.
So there exists $a$ and $b$ such that $j=a_{ab}$.
A simple arithmetic of pointers allows us to retrieve those two values expressed as integers (i.e. as distances).

Then, they are converted to binary strings: the positions of the read-only heads need a bit of pointer arithmetic to be obtained and compared, but the rest of the integer just needs to be compared bitwise.
The rest of the binary expression of the vertex encodes directly the configuration, and as all the transitions make only local changes to them, there is only a constant number of information to remember.

Every time there is a difference between the binary expression of $a$ and the binary expression of $b$, $T$ checks that the difference between them is legal regarding the transition function of $M$ ---that may be encoded in the states of $T$ or may be given as a parameter.

The transducer $T$ also have to check that $a$ and $b$ are not \enquote{phantom configurations} and that $j$ is not \enquote{too big}, i.e. does not represent a query on vertices that does not exists.
\end{proof}

\begin{corollary}
\label{cor-conl-in-ndpm}
$\cc{co-NL} \subseteq \cc{\{NDPM\}}$
\end{corollary}

\begin{proof}
Let $\pb{Pb} \in \cc{co\text{-}NL}$, there exists a non-deterministic logspace Turing machines $N$ that decides $\pb{Pb}$.
Suppose given $n \in \naturalN$, we will compose the \NDPM $M$ that solves $\stconncomp$ with the transducer $T$ that computes the graph of $N(n)$.

Every time $M$ has to read a value, it asks $T$ by letting a pointer be on the position $j$ of the value it wants to know.
There is some kind of layer of abstraction in this composition, for $M$ goes through the input tape without ever reading the actual values, but asks the values to $T$, which actually reads $n$.

We have to make sure that the $j$ of the proof of Proposition \ref{point_lemma} can be big enough: what is the size of the encoding of the graph of $N(n)$?
We encode it as being of size $2^{e \times \log(\sizeof{n})}$, i.e. we also take \enquote{phantom configurations} to be vertices.
The encoding of this \enquote{completed} graph ---for every string of size $e \times \log(\sizeof{n})$ is taken to be one of its vertex, even if it is not reachable--- is of size $\bigO{(2^{\log(\sizeof{n})})^2}$, an expression bounded by a power of $\sizeof{n}$, so we can express it.

We can suppose moreover that there is a transition between the \enquote{phantom configuration} encoded by $000\hdots001$ and the initial configuration, and that there exists a transition between any rejecting configuration and the \enquote{phantom configuration} encoded by$111\hdots 111$.
This allows us to keep the \stconncomp algorithm as is, computing only if there is no path from the vertex $1$ to the vertex $n$.

The transducer $T$ can compute the configuration graph of $N(x)$ bit-by-bit and pass it to $M$ which solves \stconncomp.
So $M\circ T(n)$ accepts iff there is no path from $1$ to a rejecting configuration in the graph of $N(n)$, i.e. iff $N(n)$ accepts.
Hence $\pb{Pb} \in \cc{NPM}$.
\end{proof}

It turns out that all \NDPMs cannot be represented as operators. Indeed, Lemma \ref{machinesperm} which establishes the equivalence between \NDPMs and operators needs an additional requirement: acyclicity. However, as we will now show, a language which is decided by a \NDPM is decided by an acyclic \NDPM.

\begin{definition}[Acyclicity]
A \NDPM $M$ is said to be \emph{acyclic} when for all $c \in C_M$ and all entry $n\in\naturalN$, $M_{c}(n)$ halts.\label{acyclic}
\end{definition}

\begin{lemma}
\label{lelemmedetransfo}
For all \NDPM $M$ that decides a set $S$ there exists an acyclic \NDPM $M'$ that decides $S$.
\end{lemma}

\begin{proof}
To prove this, we need to prove that for all $n \in \naturalN$ of size $|n|$ and $c \in C_M$ there exists a $c' \in C_{M'}$ such that if $M_c(n)$ does not halt then $M'_{c'}(n)$ rejects, and if $M_c(n)$ accepts (resp. rejects) then $M'_{c'}(n)$ accepts (resp. rejects).

We know that the number of configurations of $M$ -- with $p$ pointers -- is bounded by $|n|^p \times (3)^p \times |Q|$ that is to say bounded by $\bigO(|n|^d)$ for $d$ a constant. So we know that if $M$ does more than $\bigO(|n|^d)$ transitions, it will never halt. To obtain $M'$ we will simply add $d +1$ pointers that will behave like the hands of a clock. The first one moves forward each time we make a transition. Each time the $i$-th one has travelled through the whole input tape, the $i+1$-th one moves forward. When the last one is back on the beginning of the input tape, $M'$ rejects. It ensures us that $M'$ -- which has appart from that the same computational behaviour as $M$ -- will halt after at most $\bigO(|n|^{d+1})$ transitions.
We set $p'=p + d + 1$, and for all $\textbf{q} \in Q$, every time we had in $M$ the transition:
\[(\vec{i}, \textbf{q})\rightarrow (\vec{m}, \textbf{q}')\]
we add to $\rightarrow' \in M'$ the following set of transitions (for $p+1 \leqslant a < p'$):
\begin{align*}
(\vec{i}, \star,\hdots, \star,\textbf{q}) & \rightarrow' (\vec{m}, p_{p+1}+,\hdots, p_{p'}+, \textbf{q'})\\
(\vec{i}, 0/1,\hdots, 0/1,\textbf{q}) & \rightarrow' (\vec{m}, p_{p+1}+, \epsilon_{p+2}, \hdots, \epsilon_{p'}, \textbf{q'})\\
(\vec{i}, 0/1,\hdots, 0/1, i_a=\star, 0/1, , \hdots, 0/1,\textbf{q}) & \rightarrow' (\vec{m}, \epsilon_{p+1}, \hdots, \epsilon_{a-1},p_a+, p_{a+1}+, \epsilon_{a+2}, \hdots, \epsilon_{p'}, \textbf{q'})\\
(\vec{i},0/1, \hdots, 0/1, \star,\textbf{q}) & \rightarrow' \textbf{reject}
\end{align*}
Then, for all $c'=(\vec{i},p_{p+1},\dots,p_{p'}) \in C_{M'}$ that does not appear on the left-hand side in the previous set of transitions, we add $c' \rightarrow' \textbf{reject}$. 

For all $c=(m_1, \hdots, m_p, \textbf{q}) \in C_M$ we define $c^t=(m_1, \hdots, m_p, \star, \hdots, \star, \textbf{q}) \in C_{M'}$.

Now take a pseudo-configuration $c \in C_M$, several cases arise:
\begin{itemize}
\item If $M_c(n)$ was halting, it was in less than $\bigO(|n|^d)$ transitions so $M'_{c^t}(n)$ will have the same behavior.
\item If $M_c(n)$ was entering a loop, $M'_{c^t}(n)$ rejects after $\bigO(|n|^{d+1})$ transitions.
\end{itemize}
However, since we supposed that $M$ was deciding $S$, we know there exists a pseudo-configuration $s \in C_M$ such that for all $n \in \naturalN$, $M_{s}(n)$ halts, hence never enters a loop. As a result, by considering the pseudo-configuration $s^{t}$ we can see that $M'$ will decide the set $S$. Moreover it is clear that for all $c'\in C_{M'}$ and all $n \in \naturalN$, $M'_{c'}(n)$ always halt, so $M'$ is acyclic. 
\end{proof}

\begin{definition}
Let \cc{\{ANDPM\}} be the class of sets $S$ such that there exists an acyclic \NDPM that decides $S$.
\end{definition}

\begin{proposition}\label{co_NL_ANDPM}
\[\cc{co-NL}\subseteq \cc{\{ANDPM\}}\]
\end{proposition}

\begin{proof}
Corollary \ref{cor-conl-in-ndpm} shows that $\cc{co-NL}\subseteq \cc{\{NDPM\}}$. Moreover, it is clear that $\cc{\{ANDPM\}}\subseteq\cc{\{NDPM\}}$ and the preceding lemma shows that $\cc{\{NDPM\}}\subseteq\cc{\{ANDPM\}}$. As a consequence, we have $\cc{\{NDPM\}}=\cc{\{ANDPM\}}$ and thus $\cc{co-NL}\subseteq \cc{\{ANDPM\}}$.
\end{proof}

\section{Encoding Non-Deterministic Pointer Machines}\label{sec:encodingndpm}

\subsection{Encoding a Machine}

Our aim in this section is to prove (Lemma \ref{machinesperm}) that for any acyclic \NDPM $M$ and pseudo-configuration $c \in C_M$, there exists an observation $M_{c}^{\bullet}\in\vn{M}_{6}(\vn{G})\otimes\vn{Q}_{M}$ such that for all $N_{n}\in\vn{M}_{6}(\vn{N})$ a binary representation of $n$, $M_{c}(n)$ accepts if and only if $M_{c}^{\bullet}(N_{n} \otimes 1_{\vn{Q}_{M}})$ is nilpotent.


We will define $M_{c}^{\bullet}$ as an operator of $\vn{M}_{6}(\vn{G})\otimes\vn{Q}_{M}$, where
\begin{equation*}
\vn{Q}_{M}=\underbrace{\vn{M}_{6}(\complexN)\otimes\vn{M}_{6}(\complexN)\otimes\dots\otimes\vn{M}_{6}(\complexN)}_{p\text{ times}}\otimes\vn{M}_{s}(\complexN)
\end{equation*}
The intuition is that the $j$-th copy of $\vn{M}_{6}(\complexN)$ represents a \enquote{memory block} that contains the last value read by the $j$-th pointer. We will therefore distinguish for each copy of $\vn{M}_{6}(\complexN)$ a basis $(0o,0i,1o,1i,s,e)$ corresponding to the different values a pointer can read. The last algebra in the tensor product represents a set of states: we will distinguish a basis $Q\cup B$ where $Q$ is the set of states of the machine $M$ and $B$ is an additional set of states needed for the definition of $M_{c}^{\bullet}$. To sum up, the distinguished basis of $\vn{Q}_{M}$ considered will be denoted by tuples $(a_{1},\dots,a_{p} , \textbf{q})$. Notice that such a tuple naturally corresponds to a pseudo-configuration when $q\in Q$.

As a consequence of the tensoring of $N_{n}$ with the unit of the algebra of states, the integer is considered at the same time in every possible pseudo-configuration. As a result, the computation for $c$ a pseudo configuration represented by the sequence $M_{c}^{\bullet}(N_{n} \otimes 1_{\vn{Q}_{M}})$, $(M_{c}^{\bullet}(N_{n} \otimes 1_{\vn{Q}_{M}}))^{2}$, $\dots$ somehow simulates all the computations $M_{c}(n)$ simultaneously. However, the representation of reject cannot be done without considering an initial pseudo-configuration, something that will be explained in the next subsection.

The main difficulty is now to encode the transition relation. In order to do this, we will encode each couple $(c,t)\in\rightarrow$ by an operator $\phi_{c,t}$. The encoding of the transition relation will then correspond to the sum:
\begin{equation*}
\rightarrow^{\bullet}=\sum_{c\in C_M} \sum_{t\text{ s.t. }c \rightarrow t} \phi_{c,t}
\end{equation*}


Before explaining the encoding of basic operations, we first define the projections $\pi_{0o}$, $\pi_{0i}$, $\pi_{1o}$, $\pi_{1i}$, $\pi_{start}$, $\pi_{end}$ of $\vn{M}_{6}(\complexN)$ as the projections onto the subspace generated by the distinguished basis. We moreover define $\pi_{0\cdot}=\pi_{0i}+\pi_{0o}$ and $\pi_{1\cdot}=\pi_{1o}+\pi_{1i}$ to identify the bit currently read without considering if we come from the left (the output of the bit) or the right (the input of the bit).

For the sake of simplicity, we also define the following operators in $\vn{Q}_{M}$: if $c$ and $c'$ are respectively equal to $(a_{1},\dots,a_{p}, \textbf{q})$ and $(a'_{1},\dots,a'_{p}, \textbf{q}')$, we define the partial isometry:
\begin{equation*}
(c\rightarrow c')=(a_{1}\rightarrow a'_{1})\otimes\dots\otimes(a_{p}\rightarrow a'_{p})\otimes (\textbf{q}\rightarrow \textbf{q}')
\end{equation*}
where
\begin{equation*}
(p\rightarrow p')=\begin{array}{cc}	
	&
	\begin{array}{ccccc}
	~& ~&p&~&~
	\end{array}\\
	\begin{array}{c}
	~\\
	~\\
	p'\\
	~\\
	~
	\end{array}
	&
	\left(\begin{array}{ccccc}
	0 & \dots & 0 & \dots & 0\\
	\vdots & \ddots & \vdots& \reflectbox{$\ddots$} & \vdots\\
	0 & \dots & 1 & \dots & 0\\
\vdots & \reflectbox{$\ddots$} & \vdots & \ddots & \vdots\\
	0 & \dots & 0 & \dots & 0
	\end{array}\right)
	\end{array}
\begin{array}{c}
(p\in\{a_{1},\dots,a_{p},\textbf{q}\})\\
(p'\in\{a'_{1},\dots,a'_{p},\textbf{q}'\})\\
\end{array}
\end{equation*}
For \textbf{S} a set of states, we will use the notation $(\textbf{S} \rightarrow a'_{i})$ (denoted $(\rightarrow a'_{i})$ when $\textbf{S}$ contains all possible states) for the element that goes from any state in $\textbf{S}$ to $a'_{i}$, which is defined as $\sum_{s\in\textbf{S}} (s\rightarrow a'_{i})$.

A transition that impacts only the values stored in the subset $p_{i_{1}},\dots,p_{i_{l}}$ and the state $\textbf{q}$ will be denoted by
\begin{equation*}
([a_{i_{1}}\rightarrow a_{i_{1}}']_{i_{1}};\dots;[a_{i_{l}}\rightarrow a_{i_{l}}']_{i_{l}};\textbf{q}\rightarrow \textbf{q'})=u_{1}\otimes u_{2}\otimes\dots\otimes u_{p}\otimes (\textbf{q}\rightarrow \textbf{q'})
\end{equation*}
where $u_i=(a_{i_{j}} \rightarrow a'_{i_{j}})$ if $\exists j, i=i_{j}$, $u_{i}=\text{Id}$ elsewhere, and $\textbf{q} \rightarrow \textbf{q}'=\text{Id}$ if $\textbf{q}=\textbf{q}'$.

We are now ready to define the operators needed to encode the basic operations of the machine. Considering the von Neumann algebra $\vn{M}_{6}(\finhyp)\otimes\vn{Q}_{M}$ as $\vn{M}_{6}(\complexN)\otimes\finhyp\otimes\vn{Q}_{M}$, we will define these operators as tensor products $u\otimes v\otimes w$, where $u\in\vn{M}_{6}(\complexN)$, $v\in\vn{G}\subset\finhyp$ and $w\in\vn{Q}_{M}$.





\subsection{Basic Operations}

From now on, we consider given a machine $M$ and a pseudo-configuration $c\in C_M$.

\subsubsection{Move forward (resp. backward) a pointer, read a value and change state.}
We want to encode the action \enquote{move forward (resp. backward) the pointer $j$ when we are in the pseudo-configuration $c=(a_{1},\dots,a_{p};\textbf{q})$, read the value $a'_{j}$ stored at $\pos{p_j}$ and change the pseudo-configuration for $c'=(a_{1},\dots,a_{j-1},a'_{j},a_{j+1},\dots,a_{p};\textbf{q}')$}.
Although the operators we are going to define are all parametric in $\textbf{q}$ and $\textbf{q}'$, those parameters won't appear in their name for the sake of readability.

We first define two matrices $[\text{out}]$ and $[\text{in}]$ that will be used to keep only the values that comes next, respectively for the forward and backward move: 
\[
[\text{out}]=\left(\begin{array}{cccccc}
	1 & 1 & 1 & 1 & 1 & 1 \\
	0 & 0 & 0 & 0 & 0 & 0 \\
	1 & 1 & 1 & 1 & 1 & 1 \\
	0 & 0 & 0 & 0 & 0 & 0 \\
	1 & 1 & 1 & 1 & 1 & 1 \\
	0 & 0 & 0 & 0 & 0 & 0 \\
	\end{array}\right)
\hspace{6em}
[\text{in}]=\left(\begin{array}{cccccc}
	0 & 0 & 0 & 0 & 0 & 0 \\
	1 & 1 & 1 & 1 & 1 & 1 \\
	0 & 0 & 0 & 0 & 0 & 0 \\
	1 & 1 & 1 & 1 & 1 & 1 \\
	0 & 0 & 0 & 0 & 0 & 0 \\
	1 & 1 & 1 & 1 & 1 & 1 \\
	\end{array}\right)	
\]
The reader can refer to Definition \ref{binrep} to see that the application of one of those two matrix to the input get the desired result.

We then define the operators $\overleftarrow{m_{j}}$ and $\overrightarrow{m_{j}}$, that somehow select the $j$-th pointer thanks to the transposition $\tau_{0,j}$ that exchanges $0$ and $j$ and puts it in the right direction:
\[
\overleftarrow{m_{j}}=[\text{out}]\otimes\tau_{0,j}\otimes(\textbf{q}\rightarrow \textbf{move}_{j})\hspace{3em}
\overrightarrow{m_{j}}=[\text{in}]\otimes\tau_{0,j}\otimes(\textbf{q}\rightarrow \textbf{move}_{j})
\]
Notice that we also changed the state to $\textbf{move}_{j}$.
Finally, we define the three operators that encode the different actions the machine will make according to which of the three possible values ($0$, $1$, $\star$) the $j$-th pointer read.
Those operators come in two variants, since in case of a backward move we are going to read the output of a bit, whereas a forward move leads to the reading of the input of a bit\footnote{We consider that $\star o=\text{start}$ and $\star i=\text{end}$.}.
\begin{align*}
\overleftarrow{l_{j,0}}=\pi_{0o}\otimes\tau_{0,j}\otimes([\rightarrow \pi_{0o}]_{j};\textbf{move}_{j}\rightarrow \textbf{q}')\\
\overleftarrow{l_{j,1}}=\pi_{1o}\otimes\tau_{0,j}\otimes([\rightarrow \pi_{1o}]_{j};\textbf{move}_{j}\rightarrow \textbf{q}')\\
\overleftarrow{l_{j,\star}}=\pi_{\star o}\otimes\tau_{0,j}\otimes([\rightarrow \pi_{\star o}]_{j};\textbf{move}_{j}\rightarrow \textbf{q}')
\end{align*}
We define $\overrightarrow{l_{j,0}}$, $\overrightarrow{l_{j,1}}$ and $\overrightarrow{l_{j,\star}}$ in a similar way by substituting $i$ to $o$ in the previous equations.
Those six operators $\{ \overleftarrow{l_{j, b}}, \overrightarrow{l_{j, b}} ~|~ b \in \{0, 1, \star\}\}$ allow to move forward or backward according to the direction, when the next bit is $b$ and change the state to $\textbf{q}'$.

To sum up, we encode the backward and forward moves by:
\[
\overleftarrow{m_{j}}+ \sum_{b \in\{0, 1, \star\}} \overleftarrow{l_{j,b}}\hspace{3em} \text{ and } \hspace{3em}
\overrightarrow{m_{j}}+ \sum_{b\in\{0,1,\star\}}\overrightarrow{l_{j,b}}
\]
\subsubsection{Accept.}
The case of acceptance is especially easy: we want to stop the computation, so every transition $(a_{1},\dots,a_{n};\textbf{q})\rightarrow \textbf{accept}$ will be encoded by $0$.


\subsubsection{Reject.}\label{ssec:repreject}
We want the operator to loop to simulate the reject of the machine. Indeed, a rejection must ensure that the resulting operator $M_{c}^{\bullet}(N_{n}\otimes 1_{\vn{Q}_M})$ will not be nilpotent. A first naive attempt:
\begin{equation*}
\text{reject}_{\text{naive}}=\text{Id}_{\vn{M}_{6}(\complexN)}\otimes\text{Id}\otimes\pi_{reject}
\end{equation*}
shows that it is possible to make the computation loop, as $N_{n}^{d}\neq 0$ for all $d\in\naturalN$.
\begin{equation*}
((N_{n}\otimes 1_{\vn{Q}})\text{Id}_{\vn{M}_{6}(\complexN)}\otimes\text{Id}\otimes\pi_{reject})^{d}=(N_{n}\otimes \pi_{reject})^{d}=N_{n}^{d}\otimes \pi_{reject}
\end{equation*}
However, as $\rightarrow^{\bullet}$ is built as a sum of the basic operations, $\text{reject}_{\text{naive}}$ appears in it, and so $M^{\bullet}(N_{n}\otimes 1_{\vn{Q}_{M}})$ cannot be nilpotent\footnote{Remember that $N_{n}\otimes 1_{\vn{Q}}=N_{n}\otimes\text{Id}_{\bigotimes_{n=1}^{p}\vn{M}_{6}(\complexN)}\otimes\pi_{reject}+N_{n}\otimes\text{Id}_{\bigotimes_{n=1}^{p}\vn{M}_{6}(\complexN)}\otimes(1-\pi_{reject})$.}. This is problematic since we want this operator to be nilpotent in case of acceptance.

So we have to be a little more clever to insure the operator will loop if \emph{and only if} the operator that simulates the reject is reached. To do that, we simply make the operator go back to the chosen pseudo-configuration $c=(a_{1},\dots,a_{p};\textbf{q}_{0})$ when it reaches this operator. In this way, if reject was reached after applying the machine with a pseudo-configuration $c'$, we enforce the computation of the machine on $c$. As a consequence, if the integer was accepted by the machine in state $c$, the rejection that corresponds to a computation on $c'$ will be temporary: once rejection attained, the computation restarts with pseudo-configuration $c$ and will therefore halt accepting.

To encode this, we add two states to the machine —$\textbf{back}_{j}$ and $\textbf{move-back}_{j}$— for each $j=1,\dots,p$, and we define:
\begin{eqnarray*}
rm_{j}&=&1\otimes\tau_{0,j}\otimes(\textbf{back}_{j}\rightarrow \textbf{move-back}_{j})\\
rr_{j}&=&\pi_{0o}+\pi_{1o}\otimes\tau_{0,j}\otimes([\rightarrow \pi_{0o}+\pi_{1o}]_{j};\textbf{move-back}_{j}\rightarrow \textbf{back}_{j})\\
rc_{j}&=&\pi_{start}\otimes\tau_{0,j}\otimes([\rightarrow a_{j}]_{j};\textbf{move-back}_{j}\rightarrow \textbf{back}_{j+1})~~~(1\leqslant j<p)\\
rc_{p}&=&\pi_{start}\otimes\tau_{0,p}\otimes([\rightarrow a_{p}]_{p};\textbf{move-back}_{p}\rightarrow \textbf{q}_{0})
\end{eqnarray*}
The operator simulating the reject by making the operator loop is then defined as follows:
\[\textnormal{reject}_{c}=\left(\sum_{j=1}^p rm_j + rr_j + rc_j\right)+(\textbf{reject}\rightarrow\textbf{back}_{0})\]

\begin{definition}
Let $M$ be a pointer machine, $\rightarrow$ its transition relation and $c$ a configuration. The operator $M_{c}^{\bullet}$ is defined as:
\[M_{c}^{\bullet}=\rightarrow^{\bullet}+\textnormal{reject}_{c}\]
\end{definition}

\subsection{First Inclusions}

\begin{lemma}\label{machinesperm}
Let $M$ be an acyclic \NDPM, $c\in C_M$ and $M^{\bullet}_{c}$ the encoding we just defined. For all $n \in \naturalN$ and every binary representation $N_{n}\in\vn{M}_{6}(\vn{N}_{0})$ of $n$:
\begin{equation*}
\text{$M_c(n)$ accepts}\Leftrightarrow\text{$M^{\bullet}_{c}(N_{n}\otimes 1)$ is nilpotent}
\end{equation*}
\end{lemma}

\begin{proof}
Let us fix $n \in \naturalN$ and $N_{n}$ one of its binary representations. Considering the representation of the reject it is clear that if a branch of $M_c(n)$ rejects, the operator $M_c^{\bullet}(N_n \otimes 1)$ will not be nilpotent, so we just have to prove that if $M_c(n)$ accepts then $M_c^{\bullet}(N_n \otimes 1)$ is nilpotent. We prove its reciprocal: let's suppose $M_c^{\bullet}(N_n \otimes 1)$ is not nilpotent. In this product $N_n$ is given to the operator $M_c^{\bullet}$ that starts the simulation of the computation of $M$ with input $n$ in every possible pseudo-configuration at the same time. Since the encoding of $M$ takes in argument a pseudo-configuration $c \in C_M$, we know that there exists a $j$ such that $M_c^{\bullet}(N_n \otimes 1)\pi_j$ is the simulation of $M_c(n)$, but the computation takes place in the other projections too: for $i \neq j$ it is possible that $M_c^{\bullet}(N_n \otimes 1)\pi_i$ loops where for a $d$ $(M_c^{\bullet}(N_n \otimes 1))^d\pi_j=0$. We can correct this behavior thanks to acyclicity: if $M_c^{\bullet}(N_n \otimes 1)$ is not nilpotent it is because at some point the \textbf{reject} state has been reached. After this state of reject is reached (let's say after $r \in \naturalN$ iterations) we know that $M_c^{\bullet}(N_n \otimes 1)^{r}\pi_i$ is exactly the simulation of $M_c(n)$. If it loops again, it truly means that $M_c(n)$ rejects. So we just proved that $M_c^{\bullet}(N_n \otimes 1)$ is not nilpotent if and only if $(M_c^{\bullet}(N_n \otimes 1))^d\pi_j \neq 0$ for all $d \in \naturalN$. But it is clear that in this case $M$ with pseudo-configuration $c$ rejects the entry $n$.
\end{proof}

\begin{proposition}\label{co-NL_in_P}
\[\cc{co-NL} \subseteq \cc{\{ANDPM\}} \subseteq \{P_{+}\}\subseteq \{P_{\geqslant 0}\}\]
\end{proposition}

\begin{proof}
The first inclusion is given by Proposition \ref{co_NL_ANDPM}. By Lemma \ref{machinesperm}, we have $\cc{\{ANDPM\}}\subseteq\{P_{+}\}$ since the representation $M^{\bullet}_{c}$ of a couple $(M,c)$, where $M$ is an acyclic \NDPM and $c \in C_M$, is obviously in $P_{+}$. Moreover, since $P_{+}\subset P_{\geqslant 0}$, we have $\{P_{+}\}\subseteq\{P_{\geqslant 0}\}$.
\end{proof}

\section{Positive observations and co-NL}\label{sec:conl}

To show that $\{P_{\geqslant 0}\}$ is included in $\cc{co-NL}$, we will show that the product of a binary representation and an observation in $P_{\geqslant 0}$ is the image of a matrix by an injective morphism.
This return from the type {II}$_{1}$ hyperfinite factor to matrix algebras is necessary to prove that we can reduce the nilpotency of an operator to the nilpotency of a matrix, so that a finite machine can decide it.
This fact was used by Girard\footnote{Altough this point is not dwelled on, this statement is necessary in \cite[Proof of Theorem 12.1, p.258]{normativity}.}, but we felt it needed to be more precisely stated and proved in the following (quite technical) lemma.

\begin{lemma}\label{finiteNL}
We consider the normative pair $(\vn{N}_{0},\vn{G})$ defined in Corollary \ref{pairenormG} and denote by $\vn{K}$ the algebra $(\bigotimes_{n\geqslant 0}\finhyp)\rtimes\mathfrak{S}$. Let $N_{n}$ be a binary representation of an integer $n$ in $\vn{M}_{6}(\vn{N}_{0})$ and $\Phi\in\vn{M}_{6}(\vn{G})\otimes\vn{Q}$ be an observation in $P_{\geqslant 0}$. Then there exists an integer $k$, an injective morphism $\psi:\vn{M}_{k}(\complexN)\rightarrow\vn{K}$ and two matrices $M\in\vn{M}_{6}(\vn{M}_{k}(\complexN))$ and $\bar{\Phi}\in\vn{M}_{6}(\vn{M}_{k}(\complexN))\otimes\vn{Q}$ such that $\text{Id}\otimes\psi(M)=(N_{n}\otimes 1_{\vn{Q}})$ and $\text{Id}\otimes\psi\otimes\text{Id}_{\vn{Q}}(\bar{\Phi})=\Phi$.
\end{lemma}

\begin{proof}
We denote by $n$ the integer represented by $N_{n}$ and $R\in\vn{M}_{6(n+1)}(\complexN)$ its matricial representation. Then there exists a morphism $\theta:\vn{M}_{n+1}(\complexN)\rightarrow\finhyp$ such that $\text{Id}\otimes\theta(R)=N_{n}$ by Proposition \ref{matricerepbinaire}. Composing $\theta$ with the inclusion $\mu:\vn{M}_{n+1}(\complexN)\rightarrow\bigotimes_{n=0}^{N}\vn{M}_{n+1}(\complexN)$, $x\mapsto x\otimes 1\otimes \dots\otimes 1$, we get:
\begin{equation*}
\text{Id}\otimes(\bigotimes_{n=0}^{N}\theta(\mu(R))=\bar{N}_{n}\otimes \underbrace{1\otimes\dots\otimes 1}_{N\text{ copies}}
\end{equation*}
where $\bar{N}_{n}$ is the representation of $n$ in $\vn{M}_{6}(\complexN)\otimes\finhyp$ (recall the representation $N_{n}$ in the statement of the lemma is an element of $\vn{M}_{6}(\complexN)\otimes\vn{K}$).

Moreover, since $\Phi$ is an observation, it is contained in the subalgebra induced by the subgroup $\mathfrak{S}_{N}$ where $N$ is a fixed integer, i.e. the subalgebra of $\mathfrak{S}$ generated by $\{\lambda(\sigma)~|~\sigma\in\mathfrak{S}_{N}\}$.
We thus consider the algebra $(\bigotimes_{n=0}^{N}\vn{M}_{n+1}(\complexN))\rtimes\mathfrak{S}_{N}$. It is isomorphic to a matrix algebra $\vn{M}_{k}(\complexN)$: the algebra $\bigotimes_{n=0}^{N}\vn{M}_{n+1}(\complexN)$ can be represented as an algebra of operators acting on the Hilbert space $\complexN^{N(n+1)}$, and the crossed product $(\bigotimes_{n=0}^{N}\vn{M}_{n+1}(\complexN))\rtimes\mathfrak{S}_{N}$ is then defined as a subalgebra $\vn{I}$ of the algebra $\B{L^{2}(\mathfrak{S}_{N},\complexN^{(n+1)^{N}})}\cong\vn{M}_{(n+1)^{N}N!}(\complexN)$. We want to show that $(N_{n}\otimes 1_{\vn{Q}})$ and $\Phi$ are the images of matrices in $\vn{I}$ by an injective morphism $\psi$ which we still need to define.\\
Let us denote by $\alpha$ the action of $\mathfrak{S}_{N}$ on $\bigotimes_{n=0}^{N}\vn{M}_{n+1}(\complexN)$. By definition, $\vn{I}=(\bigotimes_{n=0}^{N}\vn{M}_{n+1}(\complexN))\rtimes\mathfrak{S}_{N}$ is generated by two families of unitaries:
\begin{itemize}
\item $\lambda_{\alpha}(\sigma)$ where $\sigma\in\mathfrak{S}_{N}$;
\item $\pi_{\alpha}(x)$ where $x$ is an element of $\bigotimes_{n=0}^{N}\vn{M}_{n+1}(\complexN)$.
\end{itemize}
We will denote by $\gamma$ the action of $\mathfrak{S}$ on $\bigotimes_{n=0}^{\infty}\finhyp$. Then $\vn{K}=(\bigotimes_{n\geqslant 0}\finhyp)\rtimes\mathfrak{S}$ is generated by the following families of unitaries:
\begin{itemize}
\item $\lambda_{\gamma}(\sigma)$ for $\sigma\in\mathfrak{S}$;
\item $\pi_{\gamma}(x)$ for $x\in\bigotimes_{n\geqslant 0}\finhyp$.
\end{itemize}
As we already recalled, $\Phi$ is an observation in $P_{\geqslant 0}$ and is thus contained in the subalgebra induced by the subgroup $\mathfrak{S}_{N}$. Moreover, $N_{n}$ is the image through $\theta$ of an element of $\vn{M}_{n+1}(\complexN)$. Denoting $\beta$ the action of $\mathfrak{S}_{N}$ on $\bigotimes_{n=0}^{N}\finhyp$, the two operators we are interested in are elements of the subalgebra $\vn{J}$ of $\vn{K}$ generated by:
\begin{itemize}
\item $\lambda_{\beta}(\sigma)$ for $\sigma\in\mathfrak{S}_{N}$;
\item $\pi_{\beta}(\bigotimes_{n=0}^{N}\theta(x))$ for $x\in\bigotimes_{n=0}^{N}\vn{M}_{n+1}(\complexN)$.
\end{itemize}
We recall that $\Phi$ is a matrix whose coefficients are finite positive linear combinations of elements $\lambda_{\gamma}(\sigma)$ where $\sigma\in \mathfrak{S}_{N}$, i.e. (denoting by $k$ the dimension of the algebra of states):
\[\Phi=(\sum_{i\in I_{a,b}}\alpha^{i}_{a,b}\lambda_{\gamma}(\sigma^{i}_{a,b}))_{1\leqslant a,b\leqslant 6k}\]
We can therefore associate to $\Phi$ the matrix $\bar{\Phi}$ defined as $\bar{\Phi}=(\sum_{i\in I_{a,b}}\alpha^{i}_{a,b}\lambda_{\alpha}(\sigma^{i}_{a,b}))_{1\leqslant a,b\leqslant 6k}$.
We will now use the theorem stating the crossed product algebra does not depend on the chosen representation (Theorem \ref{takesakithm}). The algebra $\bigotimes_{n=0}^{N} \mathfrak{M}_{n+1}(\mathbf{C})$ is represented (faithfully) by the morphism $\pi_{\beta}\circ \bigotimes_{n=0}^{\infty}\theta$. We deduce from this that there exists an isomorphism from $\vn{I}$ to the algebra generated by the unitaries $\lambda_{\beta}(\sigma)$ ($\sigma\in\mathfrak{S}_{N}$) and $\pi_{\beta}\circ\bigotimes_{n=0}^{\infty}\theta(x)$ ($x\in\bigotimes_{n=0}^{N}\vn{M}_{n+1}(\complexN)$). This isomorphism induces an injective morphism $\omega$ from $\vn{I}$ into $\vn{J}$ such that:
\begin{eqnarray*}
\omega(\pi_{\alpha}(x))&=&\pi_{\beta}(\bigotimes_{n=0}^{N}\theta(x))\\
\omega(\lambda_{\alpha}(\sigma))&=&\lambda_{\beta}(\sigma)
\end{eqnarray*}
We will denote by $\iota$ the inclusion map $\bigotimes_{n=0}^{N}\finhyp\subset\bigotimes_{n=0}^{\infty}\finhyp$ and $\upsilon$ the inclusion map $\mathfrak{S}_{N}\subset\mathfrak{S}$. We will once again use the same theorem as before, but its application is not as immediate as it was. Let us denote by $\mathfrak{S}_{N}\backslash\mathfrak{S}$ the set of the orbits of $\mathfrak{S}$ for the action of $\mathfrak{S}_{N}$ by multiplication on the left, and let us chose a representant $\bar{\tau}$ in each of these orbits. Recall the set of orbits is a partition of $\mathfrak{S}$ and that $\mathfrak{S}_{N}\times\mathfrak{S}_{N}\backslash\mathfrak{S}$ is in bijection with $\mathfrak{S}$. As a consequence, the Hilbert space $L^{2}(\mathfrak{S}_{N},L^{2}(\mathfrak{S}_{N}\backslash\mathfrak{S},\bigotimes_{n=0}^{\infty}\hil{H}))$ is unitarily equivalent to $L^{2}(\mathfrak{S},\bigotimes_{n=0}^{\infty}\hil{H})$. We will therefore represent $\bigotimes_{n=0}^{N}\finhyp$ on this Hilbert space and show this representation corresponds to $\pi_{\gamma}$. For each $x\in\bigotimes_{n=0}^{N}\finhyp$, we define $\rho(x)$ by:
\begin{equation*}
\rho(x)\xi(\bar{\tau})=\gamma(\bar{\tau}^{-1})(\iota(x))\xi(\bar{\tau})\nonumber
\end{equation*}
This representation is obviously faithful. We can then define the crossed product of this representation with the group $\mathfrak{S}_{N}$ on $L^{2}(\mathfrak{S}_{N},L^{2}(\mathfrak{S}_{N}\backslash\mathfrak{S},\bigotimes_{n=0}^{\infty}\hil{H}))$. The resulting algebra is generated by the operators (in the following, $\xi$ is an element of the Hilbert space $L^{2}(\mathfrak{S}_{N},L^{2}(\mathfrak{S}_{N}\backslash\mathfrak{S},\bigotimes_{n=0}^{\infty}\hil{H}))$):
\begin{eqnarray*}
\lambda(\nu)\xi(\bar{\tau})(\sigma)&=&\xi(\bar{\tau})(\nu^{-1}\sigma)\\
\pi(x)\xi(\bar{\tau})(\sigma)&=&\rho(\beta(\sigma^{-1})(x))\xi(\bar{\tau})(\sigma)\\
				&=&\gamma(\bar{\tau}^{-1})(\gamma(\sigma^{-1})(\iota(x)))\xi(\bar{\tau})(\sigma)\\
				&=&\gamma(\bar{\tau}^{-1}\sigma^{-1})(\iota(x)))\xi(\bar{\tau})(\sigma)\\
				&=&\gamma((\sigma\bar{\tau})^{-1})(\iota(x)))\xi(\bar{\tau})(\sigma)
\end{eqnarray*}
Through the identification of $L^{2}(\mathfrak{S}_{N},L^{2}(\mathfrak{S}_{N}\backslash\mathfrak{S},\bigotimes_{n=0}^{\infty}\hil{H}))$ and $L^{2}(\mathfrak{S},\bigotimes_{n=0}^{\infty}\hil{H}))$, we therefore get (where $\xi\in L^{2}(\mathfrak{S}_{N},L^{2}(\mathfrak{S}_{N}\backslash\mathfrak{S},\bigotimes_{n=0}^{\infty}\hil{H}))$):
\begin{eqnarray*}
\lambda(\nu)\xi(\sigma\bar{\tau})&=&\xi(\nu^{-1}\sigma\bar{\tau})\\
&=&\lambda_{\gamma}(\nu)\xi(\sigma\bar{\tau})\\
\pi(x)\xi(\sigma\bar{\tau})&=&\gamma((\sigma\bar{\tau})^{-1})(\iota(x)))\xi(\sigma\bar{\tau})\\
&=&\pi_{\gamma}(\iota(x))\xi(\sigma\bar{\tau})
\end{eqnarray*}
Applying theorem \ref{takesakithm} we finally get the existence of an injective morphism $\zeta$ from $\vn{J}$ into $\vn{K}$ such that:
\begin{eqnarray*}
\pi_{\beta}(x)&\mapsto&\pi_{\gamma}(\iota(x))\\
\lambda_{\beta}(\sigma)&\mapsto&\lambda_{\gamma}(\sigma)
\end{eqnarray*}
Figure \ref{illustr} illustrates the situation.
\begin{figure}
\centering
\begin{tikzpicture}
	\node (M) at (0,0) {$(\bigotimes_{n=0}^{N}\finhyp)\rtimes_{\beta}\mathfrak{S}_{N}$};
	\node (I) at (0,-2) {$(\bigotimes_{n\geqslant 0}\finhyp)\rtimes_{\gamma}\mathfrak{S}$};
	\node (F) at (0,2) {$(\bigotimes_{n=0}^{N}\vn{M}_{n+1}(\complexN))\rtimes_{\alpha}\mathfrak{S}_{N}$};
	\node (MG) at (4,0) {$\bigotimes_{n=0}^{N}\finhyp$};
	\node (IG) at (4,-2) {$\bigotimes_{n=0}^{\infty}\finhyp$};
	\node (FG) at (4,2) {$\bigotimes_{n=0}^{N}\vn{M}_{n+1}(\complexN)$};
	\node (MF) at (-4,0) {$\mathfrak{S}_{N}$};
	\node (IF) at (-4,-2) {$\mathfrak{S}$};
	\node (FF) at (-4,2) {$\mathfrak{S}_{N}$};
	
	\draw[->] (MG) -- (M) node [midway, above] {$\pi_{\beta}$};
	\draw[->] (IG) -- (I) node [midway, below] {$\pi_{\gamma}$};
	\draw[->] (FG) -- (F) node [midway, above] {$\pi_{\alpha}$};
	
	\draw[->] (MF) -- (M) node [midway, above] {$\lambda_{\beta}$};
	\draw[->] (IF) -- (I) node [midway, below] {$\lambda_{\gamma}$};
	\draw[->] (FF) -- (F) node [midway, above] {$\lambda_{\alpha}$};
	
	\draw[double] (FF) -- (MF) {};
	\draw[->] (MF) -- (IF) node [midway,left] {$\subset$};

	\draw[->] (MG) -- (IG) node [midway,right] {$\iota$};
	\draw[->] (M) -- (I) node [midway,left] {$\zeta$};

	\draw[->] (FG) -- (MG) node [midway,right] {$\bigotimes_{n=0}^{N}\theta$};
	\draw[->] (F) -- (M) node [midway,left] {$\omega$};
\end{tikzpicture}
\caption{Representation of the main morphisms defined in the proof of Lemma \ref{finiteNL}}\label{illustr}
\end{figure}
We now define $\psi:\vn{I}\rightarrow\vn{K}$ by $\psi=\zeta\circ\omega$. Noticing that $N_{n}=\text{Id}_{\vn{M}_{6}(\complexN)}\otimes(\pi_{\gamma}(\iota\circ\mu(\bar{N}_{n}))$, we get:
\begin{eqnarray*}
\text{Id}_{\vn{M}_{6}(\complexN)}\otimes\psi(M)&=&\text{Id}_{\vn{M}_{6}(\complexN)}\otimes\psi(\text{Id}\otimes\pi_{\alpha}(\text{Id}\otimes\mu)(R))\\
&=&\text{Id}_{\vn{M}_{6}(\complexN)}\otimes\pi_{\gamma}(\iota\circ\bigotimes_{n=0}^{N}\theta(\mu(R)))\\
&=&\text{Id}_{\vn{M}_{6}(\complexN)}\otimes\pi_{\gamma}(\iota(\bar{N}_{n}\otimes 1\otimes\dots\otimes 1))\\
&=&\text{Id}_{\vn{M}_{6}(\complexN)}\otimes\pi_{\gamma}(\iota\circ\mu(\bar{N}_{n}))\\
&=&N_{n}
\end{eqnarray*}
\begin{eqnarray*}
\text{Id}_{\vn{M}_{6}(\complexN)}\otimes\psi\otimes\text{Id}_{\vn{Q}}(\bar{\Phi})&=&(\sum_{i\in I_{a,b}}\alpha^{i}_{a,b}\psi(\lambda_{\alpha}(\sigma^{i}_{a,b})))_{1\leqslant a,b\leqslant 6k}\\
&=&(\sum_{i\in I_{a,b}}\alpha^{i}_{a,b}\lambda_{\gamma}(\sigma^{i}_{a,b}))_{1\leqslant a,b\leqslant 6k}\\
&=&\Phi
\end{eqnarray*}
The (injective) morphism $\psi$ thus satisfies all the required properties.
%
%
\end{proof}

We are now ready to prove the last inclusion to get the main theorem.

\begin{proposition}\label{P_in_co-NL}
$\{P_{\geqslant 0}\} \subseteq \cc{co-NL}$
\end{proposition}

\begin{proof}
Let $\Phi \in P_{\geqslant 0}$, $\vn{Q}$ its algebra of states and $N_{n}$ a representation of an integer $n$. By lemma \ref{finiteNL}, we know there exists a morphism $\chi$ (with $\psi$ as defined in the lemma, $\chi=\text{Id}_{\vn{M}_{6}(\complexN)}\otimes\psi\otimes\text{Id}_{\vn{Q}}$) and two matrices $M$ and $\bar{\Phi}$ such that $\chi(M\otimes1_{\vn{Q}})=N_{n}\otimes 1_{\vn{Q}}$ and $\chi(\bar{\Phi})=\Phi$. So we have $\Phi(N_{n}\otimes 1_{\vn{Q}})$ nilpotent if and only if $\bar{\Phi}(M\otimes1_{\vn{Q}})$ nilpotent. Our aim is now to prove that checking the nilpotency of this matrix is in \cc{co-NL}.\\
Our algebra is:
\[\vn{M}_{6}(\complexN)\otimes((\underbrace{\vn{M}_{n+1}(\complexN)\otimes\dots\otimes\vn{M}_{n+1}(\complexN)}_{p\text{ copies}})\rtimes\mathfrak{S}_{N})\otimes\vn{Q}\]
and we know an element of its basis will be of the form
\[(\pi,a_{0},a_{1},\dots,a_{p};\sigma;e)\]
where $\pi$ is an element of the basis $(0o,0i,1o,1i,s,e)$ of $\vn{M}_{6}(\complexN)$, $a_{i}\in\{1,\dots,k\}$ (for $i\in\{1,\dots,p\}$) are the elements of the basis chosen to represent the integer $n$, $\sigma \in \mathfrak{S}_{N}$ and $e$ is an element of a basis of $\vn{Q}$. When we apply $M\otimes 1_{\vn{Q}}$ representing the integer to an element of this basis, we obtain one and only one vector of the basis: $(\pi,a_{0},a_{1},\dots,a_{p};\sigma;e)$. When we apply to this element the observation $\bar{\Phi}$ we obtain a linear positive combination of $L\in\naturalN$ elements of the basis:
\begin{equation*}
\bar{\Phi}(\pi,a_{0},a_{1},\dots,a_{p};\sigma;e)=\sum_{i=0}^{L}\alpha_{i}(\rho,a_{\tau_{i}(0)},\dots,a_{\tau_{i}(p)};\tau_{i}\sigma;e_{i})\label{eqnilpmachines}
\end{equation*}
With a non-deterministic machine, we can follow the computation in parallel on each basis vector thus obtained. The computation can then be regarded as a tree (denoting by $b^{j}_{i}$ the elements of the basis encountered):
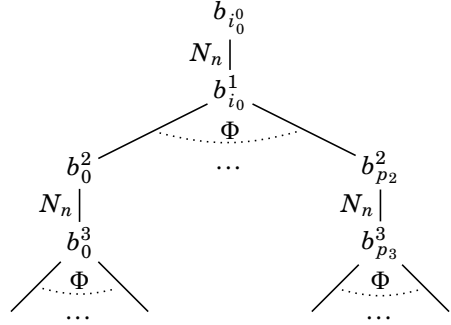
\begin{center}
\begin{tikzpicture}[x=1cm,y=0.5cm]
	\node (R) at (0,0) {$b_{i^{0}_{0}}$};
	\node (N1) at (0,-2) {$b^{1}_{i_{0}}$};
	\node (P1) at (-2,-4) {$b^{2}_{0}$};
	\node (P2) at (2,-4) {$b^{2}_{p_{2}}$};
	\node (Pmid) at (0,-4) {$\dots$};
		\node (V) at (0,-3) {$\Phi$}; 
	\node (N11) at (-2,-6) {$b^{3}_{0}$};
	\node (N22) at (2,-6) {$b^{3}_{p_{3}}$};
	\node (P11) at (-3,-8) {};
	\node (Pmid1) at (-2,-8) {$\dots$};
		\node (U) at (-2,-7) {$\Phi$}; 
	\node (P12) at (-1,-8) {};
	\node (P21) at (1,-8) {};
	\node (Pmid2) at (2,-8) {$\dots$};
		\node (T) at (2,-7) {$\Phi$}; 
	\node (P22) at (3,-8) {};
	
	\draw[-] (R) -- (N1) node [midway,left] {$N_{n}$};
	\draw[-] (N1) -- (P1) node [midway] (A) {};
	\draw[-] (N1) -- (P2) node [midway] (B) {};
	\draw[-] (P1) -- (N11) node [midway,left] {$N_{n}$};
	\draw[-] (P2) -- (N22) node [midway,left] {$N_{n}$};
	\draw[-] (N11) -- (P11) node [midway] (A1) {};
	\draw[-] (N11) -- (P12) node [midway] (B1) {};
	\draw[-] (N22) -- (P21) node [midway] (A2) {};
	\draw[-] (N22) -- (P22) node [midway] (B2) {};
	
	\draw[dotted] (A) to [bend right=15] (B) {};
	\draw[dotted] (A1) to [bend right=15] (B1) {};
	\draw[dotted] (A2) to [bend right=15] (B2) {};
\end{tikzpicture}
\end{center}
We know that $L$ and the nilpotency degree of $\bar{\Phi}(M\otimes 1_{\vn{Q}})$ are both bounded by the dimensions of the underlying space, that is to say $6(k+1)^{p}p!q$ where $q$ is the dimension of $\vn{Q}$. Since every coefficient $\alpha_{i}$ is positive, the matrix is thus nilpotent if and only if every branch of this tree is of length at most $6(k+1)^{p}p!q$.

We only have a logarithmic amount of information to store (the current basis vector), and every time a branch splits a non-deterministic transition takes place to continue the computation on every sub-branch.
\end{proof}

\begin{theorem}
\[\cc{\{ANDPM\}}=\{P_{+}\}=\{P_{\geqslant 0}\}=\cc{co-NL}\]
\end{theorem}

\begin{proof}
By combining Proposition \ref{co-NL_in_P} and Proposition \ref{P_in_co-NL}.
\end{proof}

\section{Conclusion and Perspectives}\label{sec:ccl}

This work explains the motivations and choices made by Girard when he proposed this new approach to study complexity classes. In particular, we explained how the representation of integers by matrices is an abstraction of sequent calculus proofs of the type of binary lists in \textbf{ELL}, and how using the hyperfinite factor allows to overcome the lack of uniformity of the matrix representation. We then introduced a notion of normative pair which differs from the one introduced by Girard and showed how the crossed product construction can be used to define such pairs. Going from an interaction based on the determinant to one relying on nilpotency allows to consider a larger class of groups in the construction based on the crossed product. Moreover, even if the two definitions are equivalent in some cases, such as the one considered in this paper, they differ in some others. 

We then introduced non-deterministic pointer machines as a technical tool to show that $\cc{co-NL}\subseteq\{P_{+}\}$. The proof of this inclusion, which was only sketched in Girard's paper, helps to get more insights on how computation is represented by operators. Moreover, it gives a new characterization of \cc{co-NL} in term of machines. We then proved that $\{P_{\geqslant 0}\}\subseteq\cc{co-NL}$ following the proof given by Girard \cite{normativity}, providing a proper statement and a proof of the key technical result that was not provided by Girard.
Of course, we could have used the famous result which states that \pb{REACHABILITYComp}, is in \cc{NL} \cite{immerman1988nondeterministic}, to prove that we also characterized \cc{NL}, but we hope to get a different proof of this closure by complementation with our tools.

We believe that this new approach of complexity can be used to characterize other complexity classes. Two different possibilities should be considered: changing the normative pair, and changing the set of observations. As we showed, one could define a normative pair from a group action by using the crossed product construction. However, obtaining new results in this way requires to overcome the difficulty of finding appropriate groups.

The second possibility, which seems at the time less complicated, would be to consider other sets of observations for the same normative pair. For instance, one could define the set of observations whose coefficients are unitaries induced by group elements and whose norm is equal to $1$ (so that there are at most one non-zero coefficient in each column). Denoting this set by $P_{1}$, we can easily adapt the proof of Proposition \ref{P_in_co-NL} to show that $\{P_{1}\}\subseteq \textbf{L}$. However, the question of whether the corresponding class $\{P_{1}\}$ is equal or strictly included in $\cc{L}$, and its eventual relations to PURPLE \cite{pointer08}, still need to be answered. 

%


\bibliography{biblio}

@string{Takesaki="Takesaki, Masamichi"}

@string{Seiller="Seiller, Thomas"}

@string{DalLago="Dal Lago, Ugo"}

@string{Aubert="Aubert, Clément"}

@string{Girard="Girard, Jean-Yves"}

@string{Hofmann="Hofmann, Martin"}

@string{Schopp="Ulrich Schöpp"}

@string{TheorComputSci="Theoretical Computer Science"}

@string{MS="Mathematica Scandinavia"}

@string{IaC="Information and Computation"}

@string{LNCS="Lecture Notes in Computer Science"}

@string{FI="Fundamenta Informaticae"}

@string{LICS={LICS}}

@string{fossacs={FoSSaCS}}

@string{DICE={DICE}}

@string{ieeecs={IEEE Computer Society}}

@string{Immerman={Immerman, Neil}}

@string{LMCS="Logical Methods in Computer Science"}

@string{coco="CoCo"}

@BOOK{arora2009,
title={Computational complexity: a modern approach},
publisher={Cambridge University Press},
year={2009},
author={Arora, Sanjeev and Barak, Boaz},
volume={1}
}

@inproceedings{aubert11,
author=Aubert,
title={Sublogarithmic uniform Boolean proof nets},
booktitle=DICE,
year={2011},
pages={15-27},
doi={10.4204/EPTCS.75.2},
crossref={dice12}
}

@proceedings{dice12,
editor={Marion, Jean-Yves},
title={Proceedings Second Workshop on Developments in Implicit Computational Complexity},
booktitle=DICE,
series={EPTCS},
volume={75},
year={2011},
doi={10.4204/EPTCS.75},
}

@article{baillotpedicini,
title={Elementary Complexity and Geometry of Interaction},
author={Baillot, Patrick and Pedicini, Marco},
journal=FI,
volume={45},
number={1-2},
pages={1--31},
year={2001},
publisher={IOS Press},
doi={10.1007/3-540-48959-2_4}
}

@book{conway,
Author={Conway, John B.},
Publisher={Springer},
Title={A course in functional analysis},
Year={1990}
}

@inproceedings{lago05,
author=DalLago,
title={The Geometry of Linear Higher-Order Recursion},
booktitle=LICS,
year={2005},
pages={366--375},
crossref={lics2005}, 
doi={10.1109/LICS.2005.52}
}

@article{dal09,
author=DalLago#{ and }#Hofmann,
title={Bounded Linear Logic, Revisited},
journal=LMCS,
year={2010},
volume={6},
number={4},
pages={1--31},
doi={10.2168/LMCS-6(4:7)2010}
}

@article{danos01,
Author={Danos, Vincent and Joinet, Jean-Baptiste},
Journal=IaC,
Title={Linear Logic \& Elementary Time},
Volume={183},
year={2003},
pages={123--137},
number={1},
publisher={Elsevier},
doi={10.1016/S0890-5401(03)00010-5}
}

@article{goi1,
Author=Girard,
journal={Studies in Logic and the Foundations of Mathematics},
Title={Geometry of Interaction $\text{I}$: Interpretation of System F},
Year={1989},
volume={127},
pages={221--260},
publisher={Elsevier},
doi={10.1016/S0049-237X(08)70271-4}
}

@INPROCEEDINGS{goi3,
author=Girard,
title={{T}owards a geometry of interaction},
booktitle={Proceedings of the AMS Conference on Categories, Logic and Computer Science},
year=1989,
pages={69--108},
doi={10.1090/conm/092/1003197}
}

@ARTICLE{goi5,
author=Girard,
title={{G}eometry of {I}nteraction {V}: {L}ogic in the {H}yperfinite {F}actor},
journal=TheorComputSci,
year={2011},
volume={412},
pages={1860--1883},
number={20},
doi={10.1016/j.tcs.2010.12.016},
publisher={Elsevier},
}

@incollection{normativity,
Author=Girard,
title={Normativity in Logic},
pages={243--263},
doi={10.1007/978-94-007-4435-6_12},
crossref={epistemo_vs_onto2012}
}

@book{epistemo_vs_onto2012,
editor={Dybjer, Peter and Lindström, Sten and Palmgren, Erik and Sundholm, Göran},
title={Epistemology versus Ontology - Essays on the Philosophy and Foundations of Mathematics in Honour of Per Martin-Löf},
booktitle={Epistemology versus Ontology},
publisher={Springer},
series={Logic, Epistemology, and the Unity of Science},
volume={27},
year={2012},
isbn={978-94-007-4434-9, 978-94-007-4435-6},
doi={10.1007/978-94-007-4435-6}
}

@article{haagerup,
Author={Haagerup, Uffe},
Journal=MS,
Number={271-283},
Title={The Standard Form of von Neumann Algebras},
Volume={37},
Year={1975}
}

@inproceedings{Hofmann2013,
author=Hofmann #{ and Ramyaa, Ramyaa and }#Schopp,
title={Pure Pointer Programs and Tree Isomorphism},
booktitle=fossacs,
year={2013},
pages={321-336},
doi={10.1007/978-3-642-37075-5_21},
crossref={fossacs2013}
}

@proceedings{fossacs2013,
editor={Pfenning, Frank},
title={Foundations of Software Science and Computation Structures - 16th International Conference, FOSSACS 2013, Held as Part of the European Joint Conferences on Theory and Practice of Software, ETAPS 2013, Rome, Italy, March 16-24, 2013. Proceedings},
booktitle=fossacs,
publisher={Springer},
series=LNCS,
volume={7794},
year={2013},
isbn={978-3-642-37074-8},
doi={10.1007/978-3-642-37075-5}
}

@inproceedings{immerman1988nondeterministic,
author=Immerman,
title={Nondeterministic space is closed under complementation},
year={1988},
pages={112--115},
crossref={coco1988},
doi={10.1109/SCT.1988.5270},
}

@article{lafont04,
title={Soft linear logic and polynomial time},
author={Lafont, Yves},
journal=TCS,
volume={318},
number={1},
pages={163--180},
year={2004},
publisher={Elsevier}
}

@book{murphy,
Address={Boston, MA},
Author={Murphy, Gerard J.},
Publisher={Academic Press Inc.},
Title={{C$^{\ast}$}-algebras and operator theory},
Year={1990},
}

@article{rosenberg66,
title={On multi-head finite automata},
author={Rosenberg, A.L.},
journal={IBM Journal of Research and Development},
volume={10},
number={5},
pages={388--394},
year={1966},
publisher={IBM},
doi={10.1147/rd.105.0388}
}

@inproceedings{schopp07,
author=Schopp, 
title={Stratified Bounded Affine Logic for Logarithmic Space},
booktitle=LICS,
year={2007},
pages={411-420},
bibsource={DBLP, http://dblp.uni-trier.de},
crossref={lics2007},
doi={10.1109/LICS.2007.45}
}

@proceedings{lics2007,
title={22nd {IEEE} Symposium on Logic in Computer Science (LICS 2007), 10–12 {J}uly 2007, Wroclaw, Poland, Proceedings},
year={2007},
publisher={IEEE Computer Society},
booktitle=LICS
}

@inproceedings{pointer08,
author=Hofmann#{ and }#Schopp,
title={Pointer Programs and Undirected Reachability},
booktitle=LICS,
year={2009},
pages={133--142},
doi={10.1109/LICS.2009.41},
crossref={lics2009},
}

@article{seiller-goiadd,
Author=Seiller,
Title={Interaction Graphs: Additives},
Year={2012},
journal={Arxiv preprint},
archivePrefix={arXiv},
volume={abs/1205.6557},
eprint={1205.6557},
primaryClass={cs.LO}
}

@article{seiller-goim,
Author=Seiller,
Title={Interaction Graphs: Multiplicatives},
Journal={Annals of Pure and Applied Logic},
Pages={1808-1837},
Volume={163},
Month={December},
Year={2012},
DOI={10.1016/j.apal.2012.04.005}
}

@phdthesis{seiller-phd,
Author=Seiller,
School={Universit{\'e} de la M{\'e}diterran{\'e}e},
Title={Logique dans le facteur hyperfini : g{\'e}ometrie de l'interaction et complexit{\'e}},
Url={http://tel.archives-ouvertes.fr/tel-00768403/},
Year=2012
}

@book{Takesaki3,
Author=Takesaki,
Publisher={Springer},
Series={Encyclopedia of Mathematical Sciences},
Title={Theory of Operator Algebras 3},
Volume={127},
Year={2003},
}

@book{Takesaki2,
Author=Takesaki,
Publisher={Springer},
Series={Encyclopedia of Mathematical Sciences},
Title={Theory of Operator Algebras 2},
Volume={125},
Year={2003},
}

@book{Takesaki1,
Author=Takesaki,
Publisher={Springer},
Series={Encyclopedia of Mathematical Sciences},
Title={Theory of Operator Algebras 1},
Volume={124},
Year={2001},
}

@InProceedings{terui04,
AUTHOR="Terui, Kazushige",
TITLE="{Proof Nets and Boolean Circuits}",
booktitle=LICS,
year={2004},
pages={182--191},
crossref={lics2004},
doi={10.1109/LICS.2004.1319612}
}

@proceedings{lics2004,
title={19th {IEEE} Symposium on Logic in Computer Science (LICS 2004), 14–17 {J}uly 2004, Turku, Finland, Proceedings},
year={2004},
publisher={IEEE Computer Society},
booktitle=LICS,
isbn={0-7695-2192-4}
}

@proceedings{lics2009,
title={Proceedings of the 24th Annual IEEE Symposium on Logic in Computer Science, LICS 2009, 11–14 August 2009, Los Angeles, CA, USA},
booktitle=LICS,
publisher=ieeecs,
year={2009},
isbn={978-0-7695-3746-7}
}

@proceedings{coco1988,
title={Proceedings: Third Annual Structure in Complexity Theory Conference, Georgetown University, Washington, D. C., USA, June 14-17, 1988},
booktitle=coco,
publisher=ieeecs,
year={1988},
isbn={0-8186-0794-7},
ee={http://ieeexplore.ieee.org/xpl/mostRecentIssue.jsp?punumber=209}
}

@proceedings{lics2005,
title={20th IEEE Symposium on Logic in Computer Science (LICS 2005), 26–29 June 2005, Chicago, {IL}, {USA}, Proceedings},
booktitle=LICS,
publisher=ieeecs,
year={2005},
isbn={0-7695-2266-1}
}

@article{Szelepcsenyi1987,
author={Szelepcsényi, Róbert},
title={The method of focing for nondeterministic automata},
journal={Bulletin of the EATCS},
volume={33},
year={1987},
pages={96-99}
}
\end{document}